\documentclass[11pt,twoside]{article}

\usepackage{baaa}

\usepackage{graphicx}
\usepackage{subfigure}
\usepackage{amssymb}
\usepackage[spanish,activeacute,english]{babel}
\usepackage[latin1]{inputenc}
\usepackage{verbatim}
\usepackage{amsmath}
\usepackage{amsfonts}
\usepackage{amssymb}
\usepackage[colorlinks=true,dvips]{hyperref}
\usepackage{natbib}
\usepackage{lscape}

\begin{document}

\vskip 1.0cm
\markboth{G. Gunthardt et al.}%
{ }

\title{Kinematics of M51-type interacting galaxies}

\author{G. I. G\"unthardt}
\affil{Observatorio Astron\'omico, Universidad Nacional de C\'ordoba, Argentina. \\(guillermo.gunthardt@unc.edu.ar)}

\vskip 0.8cm
\author{R. J. D\'iaz}
\affil{Gemini Observatory, AURA, USA.}
\affil{ICATE, CONICET, Argentina.}

\vskip 0.8cm

\author{M. P. Ag\"uero}
\affil{Observatorio Astron\'omico, Universidad Nacional de C\'ordoba and CONICET, Argentina.}

\begin{abstract} We present a kinematic catalogue for 21 M51-type galaxies. It consists of radial velocity distributions observed with long slit spectroscopy along different position angles, for both the main and satellite components. We detect deviations from circular motion in most of the main galaxies of each pair, due to the gravitational perturbation produced by the satellite galaxy. However some systems do not show significant distortions in their radial velocity curves. We found some differences between the directions of the photometric and kinematic major axes in the main galaxies with a bar subsystem. The Tully-Fisher relation in the B-band and Ks-band for the present sample of M51-type systems is flatter when compared with isolated galaxies. Using the radial velocity data set, we built a synthetic normalized radial velocity distribution, as a reference for future modeling of these peculiar systems. The synthetic rotation curve, representing the typical rotation curve of the main galaxy in an M51-type pair, is near to solid body-like inside 4 kpc, and then is nearly flat within the radial range 5-15 kpc. The relative position angles between the main galaxy major axis and the companion location, as well as the velocity difference amplitude, indicate that the orbital motion of the satellite has a large projection on the main galaxy equatorial plane. In addition, the radial velocity differences between the two galaxies indicate that the satellite orbital motion is within the range of amplitudes of the main galaxy rotation curve and all the M51-type systems studied here except for one, are gravitationally bounded.
\end{abstract}

\keywords{galaxies: interactions -- galaxies: kinematics and dynamics }

\section{Introduction}
\label{Int}

\noindent M51-type systems are interacting galaxies composed of a larger main galaxy and a smaller or satellite galaxy near to or at the end of a tidal arm developed by the larger component. Vorontsov-Velyaminov (1975) considered a subset of 160 of this type of interacting galaxies and many of these objects have been included in his \textit{``Atlas of Interacting Galaxies"} (1977). Halton Arp also included 54 of these systems in the \textit{``Atlas of Peculiar Galaxies"} (1966) and they are considered as a category of interaction in the \textit{``Catalogue of Southern Peculiar Galaxies and Associations"} of Arp $\&$ Madore (1987). Jokim\"aki et al. (2008) built a catalogue of more than 200  apparently interacting galaxy pairs of the M51 class, from visual identification of IRSA archives, and found that a low number of the main galaxies in M51 systems are early type spirals and barred spirals and that about 70\% of the main galaxies in M51 systems are 2-armed spirals. \par
M51-type galaxies are a subset of the class of unequal mass interacting galaxy pairs, for which there is some statistical evidence based on large surveys for enhanced star formation relative to normal galaxies, particularly in the smaller of the galaxies in the pair (e.g., Woods $\&$ Geller 2007; Ellison et al. 2008).

Ellison et al. (2008), from a sample of 1716 galaxies, members of galaxy pairs from Sloan Digital Sky Survey Data Release 4, found that the star-forming rate (SFR) enhancement can be detected for a sample of galaxy pairs whose masses are within a factor of 10 of each other.

According to Smith et al. (2007), a M51-like subset of Arp galaxies also shows possible differences from the spirals in mid-infrared colors; however, the M51 sample size is small, so these results are tentative. They obtain for eight M51-like systems, redder [8]\,-\,[24], [3.6]\,-\,[24], [5.8]\,-\,[8] colors, when compared to spirals, and therefore enhanced star formation. The estimated median SFR (SFR derived from the 24\,$\mu$m luminosity, by using the calibration from Calzetti et al. 2005) for their small sample of M51-type galaxies resulted in 3.3 \textit{M$_{\odot}$} yr$^{-1}$, while the median for the SFR of the spirals of their control sample is of 0.9 \textit{M$_{\odot}$} yr$^{-1}$. For comparison, these authors also find for their sample of 35 tidally distorted premerger interacting galaxy pairs, which were selected from the Arp Atlas, a SFR of $\sim$1.7 \textit{M$_{\odot}$} yr$^{-1}$. For the mentioned sample of 35 Arp systems, they derived a SFR of $\sim$2.6 \textit{M$_{\odot}$} yr$^{-1}$ from the 8$-$1000 $\mu$m total infrared luminosity as a star-forming indicator. \par
Laurikainen, Salo \& Aparicio (1998) obtained deep broad-band BVRI photometry of close interacting galaxy pairs and found that 9 of 13 M51-type pairs showed enhanced star formation in the central regions of the companions and only one pair in the nucleus of the main galaxy.  \par
Klimanov $\&$ Reshetnikov (2001), made a final selection of 32 M51-like objects from the list of Vorontsov-Velyaminov and performed a statistical study using photometric information from B-band of the \textit{``Digitized Sky Survey"} (DSS) and far-infrared from the IRAS satellite. These authors conclude that there is an enhancement of the star-forming activity with respect to isolated galaxies. In fact, from IRAS data, they obtained an average SFR for the main galaxies of M51-type systems, of 9 \textit{M$_{\odot}$}/yr. They compare these results with those of Bushouse (1987) for isolated galaxies, finding that the average SFR for this sample of M51-type galaxies is about 7 times that of isolated galaxies. In summary, broad band photometry surveys indicate that M51 galaxies are places of active star formation.
\\
\\ From the kinematic perspective, which is our main goal in the present work, we can assert that M51-like galaxies are not well studied systems. The kinematic data for this type of objects is scarce, and generally, for many of them, only the systemic velocity of the large galaxy is available. A long-slit spectroscopic kinematic study was previously performed by Klimanov, Reshetnikov $\&$ Burenkov (2001) who obtained kinematic data for 13 M51-type systems and considered radial velocity determinations from the literature for other 8 systems. From the mentioned data, Reshetnikov $\&$ Klimanov (2003) found that moderately massive dark halos surround bright spiral galaxies and that the Tully-Fisher relation for M51-like galaxies was flatter when compared with local field galaxies.
 Other kinematic studies of this class of objects have been carried out from Fabry-Perot interferometry observations. Rampazzo et al. (2005) have mapped, by using the H$\alpha$ emission line, the warm gas distribution and the velocity fields of two M51-like systems, Arp\,70 and Arp\,74. They detected gas motions following the elongated arm/tail of Arp\,70b, while in the fainter member, Arp\,70a, the gas distribution is off-centered with respect to the stellar isophotes, suggesting to the authors that it could be due to an external acquisition. They also detected non-circular motions in the velocity field of the main galaxy of Arp\,74, Arp\,74a. Fuentes-Carrera et al. (2007) studied the kinematics and dynamics of the M51-type pair NGC\,3893/96, also from Fabry-Perot interferometry, and detected non-circular motions on the velocity fields of both galaxies, probably due to the encounter.  \\
 In this paper we aim to determine how common the kinematic perturbations reported before are, as well as to characterize the typical rotation curve and range of masses, using detailed observations of a relatively large sample of M51 candidate objects.

\section{Observations}

In what follows, we present a kinematic catalogue for 21 M51-type interacting systems (Table\,1) corresponding to spectroscopic observations performed at \textit{``Complejo Astron\'omico El Leoncito"}, San Juan, Argentina, in six observing runs from 2005 to 2007 (April and October). We used the 2.15 m telescope and a REOSC spectrograph, in long-slit mode. A 1200 l/mm grating was used, and the resulting spectral resolution was of about 3 \AA{}. The seeing during observations fluctuated between 2$\arcsec$ and 3$\arcsec$, so the slit width was opened in 250 $\mu$m or 300 $\mu$m, according to the observing conditions. The wavelength range was between $\sim$ 6200 and 7000 \AA{}. An example spectrum is shown in Figure\,3. Generally, the slit was positioned on the nuclear region of both components of the M51-type systems and along the photometric main axis of both galaxies. For some objects, additional directions were chosen. Data reduction of the spectra was made employing the standard methods of the IRAF (Image Reduction and Analysis Facility)\footnote{http://iraf.noao.edu/ IRAF is distributed by the National Optical Astronomy Observatory, which is operated by the Association of Universities for Research in Astronomy (AURA) under a cooperative agreement with the National Science Foundation.} reduction package. 
 Radial velocities were obtained fitting a Gaussian profile to the H$\alpha$ emission line, while uncertainties were determined from the empirical expression for $\sigma$ by Keel (2004). An uncertainty of 2\,$\sigma$ was assigned for each measured radial velocity.

\section{Results}

In Table\,1 we listed the selected objects, with their absolute equatorial coordinates and B-band magnitudes, which have been extracted from LEDA (Lyon-Meudon Extragalactic Database){\footnote{http://leda.univ-lyon1.fr/}}. The B-band images of these objects are shown in Figure\,1. We have measured Ks-band aperture magnitudes for both main and satellite galaxies from the corresponding calibrated 2MASS (The Two Micron All Sky Survey) images, since generally there is no information for satellite galaxies in the 2MASS photometric catalogues. Details of the photometry procedure may be found in G\"unthardt (2009). In Table\,1 we include Ks-band apparent magnitudes for main and satellite galaxies of the sample and in Figure\,2 we show histograms corresponding to the Ks-band (2MASS) luminosity for main galaxies and Ks luminosity ratios between satellite and main galaxies of this sample. Ks-band magnitudes have been corrected for Galactic extinction (Schlafly \& Finkbeiner 2011), while internal extinction have been estimated by following the expressions of Masters, Giovanelli \& Haynes (2003) (see their table 5). For the k-correction (\textit{k$_{K}$}) we considered a linear fit for the low-redshift end of the models of Poggianti (1997), where  \textit{k$_{K}$} = -1,52 z. The cosmological constant has been adopted as \textit{H$_{0}$} = 75 km s$^{-1}$ Mpc$^{-1}$ and the K-band luminosities have been determined using \textit{L$_{K}$} = 10\,$^{-0.4(\textit{\scriptsize{M$_{K}$}}\,-\,3.28)}$.

 In Figure\,1, we plot the B-band ``Digitized Sky Survey" (DSS) images and the positions of the slit. Figure\,4 shows, for the different observed position angles, the heliocentric radial velocity distributions (not corrected for the inclination of the disks with respect to the direction of the sight). In some cases, the observed radial velocity gradients are drawn.
In Table\,2 we present the systemic velocities corresponding to main and satellite components, and in Figure\,2 we show an histogram of the  systemic radial velocity differences between main and satellite galaxies.
 For each observed system presented in this catalogue, a qualitative kinematic analysis was done. We also present a comparison between kinematic and photometric major axis. The Tully-Fisher relation (Figure\,7) is plotted with radial velocity data derived from this work and B-band and Ks-band magnitudes. Finally synthetic radial velocity curves are derived for the main galaxies.

\subsection{General morphological characteristics of the sample}

We present a brief description of the morphology of the systems searching for common characteristics, by considering the projected separation between components, development of the tidal arms and inclination of the disks of the components.\\
Some of the studied interacting systems have a very well developed tidal arm while their satellites have a disk-like structure and are nearly edge-on. Examples corresponding to this group are VV\,410, NGC\,633, AM\,0459-340 and AM\,2058-381. Other systems present a well-developed tidal arm, and with satellites having E type or irregular morphologies, like AM\,0639-582 and AM\,0430-285.
The opposite case is represented by some pairs which are apparently very near to each other, like NGC\,4188, VV\,452, AM\,1325-274 and AM\,0403-604. It is also possible to find similar configurations in AM\,1304-333 and AM\,0327-285, but now with satellites having a considerable mass in relation to the mass of their main galaxies. There are intermediate cases in which a not-well developed tidal arm is seen, with an apparent lack of contact with the satellite: AM\,1416-262 and ARP\,54. The main galaxy in Arp\,54 is a typical ocular-shaped galaxy, that is, a galaxy that contains a central transient eye-shaped structure. This may be the result of an encounter with a smaller galaxy (for more details, see Thomasson 2004).
Other subgroup include systems where their tidal arms are rather diffuse, like AM\,2103-332 and VV\,350. In regard to nuclear activity, the main galaxy of AM\,0403-604 is the only Seyfert 2 type galaxy of this kinematic sample (G\"unthardt 2009) having a very near edge-on satellite and, in addition to the tidal arm, a bridge is seen connecting the galaxies. AM\,1416-262 has a main galaxy which is the only Seyfert 1 of the sample.

\subsection{Analysis per object}

\medskip
\textbf{\hspace{7.5mm}NGC\,633}

\smallskip
In the DSS picture, the main galaxy presents two well-developed arms. Both arms end in a diffuse fashion, near the satellite galaxy, so it would be possible to consider this object as a not typical M51-type system, taking into account that it is the shorter arm that is closer in projection to the satellite galaxy. Apparently, the satellite would be the cause of the longer extension that presents the longer tidal arm. This tidal arm seems to begin in the south-western part of the disk of the main galaxy, although it is difficult to locate exactly in which part of the disk it originates, because of its very diffuse structure. The longer arm could have been disturbed by the passage of the satellite galaxy, and actually, the smaller galaxy would almost coincide with the position of the shorter, and less perturbed arm, which originates in the eastern side of the disk main galaxy.
From H$\alpha$ images obtained by Dopita et al. (2002), we observed that the emission concentrates in the nuclear regions of both galaxies and that the distribution of the H$\alpha$ emission in the satellite is perpendicular to its major axis, as was noticed by Dopita et al. (2002). As these authors pointed out, this could be due to the presence of outflow.
In what follows, we analyze the radial velocity data for different position angles (PA).\\
PA\,136$^{\circ}$: The observed gradient in the central kpc (7") is of 55\,km\,s$^{-1}$\,kpc$^{-1}$. If we take into account all the rigid body extension in the curve, the gradient is of about 30\,km\,s$^{-1}$\,kpc$^{-1}$. The general aspect of the curve is rather normal.\\
PA\,90$^{\circ}$: We noted some disturbances in this position angle. From -5\,kpc (-14") until reaching the position near the center, we measure an observed gradient of 10\,km\,s$^{-1}$\,kpc$^{-1}$ , while from the center to 3\,kpc (9") is about 30\,km\,s$^{-1}$\,kpc$^{-1}$. Beyond the central 10" the radial velocity gradients decrease and the variations reach 75\,km\,s$^{-1}$ which may be associated with the arms, although these significant amplitude variations may be due to the gravitational perturbation of the satellite. We also found asymmetries in the radial velocity distributions along PA\,=\,53\,$^{\circ}$.\\
For this system, we have obtained spectra for three directions, so it was possible to determine its main kinematic axis, which is coincident with its main photometric axis.
Generally, some mild asymmetries and distortions are observed in the main galaxy radial velocity distribution.\\
The radial velocity distributions along the satellite main axis is rigid body-like in practically the whole extension.

\medskip
\textbf{ARP\,54}

\smallskip
The main galaxy is classified as SABc and presents an ``ocular" morphology, a denomination that makes reference to the apparent shape of the ring of the galaxy and the nuclear region. The difference in radial velocity between both components is of $\sim$\,60\,km\,s$^{-1}$ (the satellite radial velocity had no previous determinations and have been performed in CASLEO with the same instrumental configuration described above, but with a 300 lines\,mm$^{-1}$ grating). The arm that would connect with the satellite abruptly interrupts, in both 2MASS and in DSS images. The radial velocity distribution along PA\,90$^{\circ}$ for the main galaxy was measured from very faint H$\alpha$ emission in its central 6", so it was not possible to derive kinematic information for the central region. The curve does not present significant distortions: the variations observed beyond 10" to the western direction correspond to the arm  positions (and to the position of an emission region placed between the disk and the western arm). Only a slight asymmetry is noted in PA\,0$^{\circ}$: the south region of the curve is almost constant and it rises 50\,km\,s$^{-1}$ in the northern part, precisely from the north where the tidal arm begins. This asymmetry is the only sign of perturbation due to the satellite.

\medskip
\textbf{AM\,0327-285}

\smallskip
This system presents a relatively small separation between the components. In the direction that connects both objects, the distribution has an observed maximum of 250\,km\,s$^{-1}$ with some slight asymmetries, as a rise in the radial velocity toward the direction of the satellite. The perturbation perhaps is more notorious along PA\,30$^{\circ}$, where the peak of the continuum emission coincides with the position that belongs to the maximum in radial velocity. From this position, to the SW, the curve decreases about 100\,km\,s$^{-1}$. For PA\,90$^{\circ}$, we have found variations of about 100\,km\,s$^{-1}$, which spatially correspond to the position of the spiral arms.

\medskip
\textbf{AM\,0403-604}

\smallskip
In this system, besides its M51-type nature, a bridge is apparently connecting both galaxies. The main galaxy has a nuclear emission corresponding to Seyfert 2 type (as determined from spectrophotometric observations performed at CASLEO in October 2007). The velocity difference between main and satellite galaxy, which had no previous determinations, is rather high, being of 490\,km\,s$^{-1}$. It is suggestive that one of two active objects of the sample presents a direct connection (bridge) between the central region of the galaxy and its companion.

\medskip
\textbf{AM\,0430-285}

\smallskip
The morphology of this system is similar to NGC\,646, an M51 system which is not included in this work, but observed in CASLEO for a spectrophotometric study. A small condensation joined by a faint bridge/arm with the satellite galaxy is observed. This object did not have previous radial velocity measurements, the present work confirms that it constitutes a physical pair with the main galaxy. The direction which connects the center of main and satellite galaxies, coincides with the major axis of the satellite, which presents rotation, with an amplitude of 60\,km\,s$^{-1}$. The orbit of the system is prograde, and the radial velocity distribution of the main galaxy is asymmetric with respect to the position of the continuum emission peak. For the direction along the bar, which is nearly coincident with its photometric major axis, the velocity distribution presents pronounced breaks. The variations outside the rigid-body zone, reach 70\,km\,s$^{-1}$ NE off the center. If we take into account that the semi-amplitude of the rotation curve is of 140\,km\,s$^{-1}$, results are very significant.

\medskip
\textbf{AM\,0458-250}

\smallskip
The main galaxy of this system is large (according to LEDA data, its 25\,mag/arcsec$^{2}$ diameter in B-band is of 1.15' or about 50\,kpc). The main galaxy is a massive spiral (the mass may be estimated from the velocity distribution, in a Keplerian approximation). Its satellite had no previous velocity determination and they are confirmed as a physical pair though no significant distortions are seen in the velocity distribution.


\medskip
\textbf{AM\,0459-340}
\smallskip

The heliocentric radial velocity curve, along the major axis of the main galaxy, presents high asymmetry respect to the position of the nuclear region. The observed radial velocity gradient of the NE branch is of 35\,km\,s$^{-1}$\,kpc$^{-1}$, reaching a maximum 10" from the center, and with a difference of 80\,km\,s$^{-1}$ with respect to that of the central part. Here we find that the satellite forms a physical pair with the main galaxy. The satellite depicts a rigid body curve, in most of its extension. The radial velocity gradient notably increments in the region of the bridge that connects with the Eastern condensations, so it would be possible that this small object was not previously linked with the satellite galaxy.

\medskip
\textbf{ESO\,362-IG01}
\smallskip

This system, 3' far from AM\,0459-340, was not included previously in the M51 type category of interacting systems. No significant distortions are observed although an asymmetry in the H$\alpha$ emission can be observed. This emission is more intense in the opposite side to the satellite and is the only part of the curve with a flat distribution (the outermost regions of the NE branch of the curve). For the other parts, the curve presents a rigid body distribution. The satellite had no previous radial velocity measurements, so we confirm the physical link between both components.

\medskip
\textbf{AM\,0639-582}
\smallskip

This system has a long tidal tail, which extends from the spiral main galaxy to an irregular galaxy. The radial velocity distribution along the main galaxy major axis is apparently normal, but there is a well-marked difference in the observed gradients, being the gradient more pronounced in the satellite direction, between 4" north from the center and 3" to the south, while the gradient is lower between 4" and 18" to the south. Observations along the centers of both galaxies were also made, determining that the orbit is prograde. The satellite radial velocity is similar to the velocity that corresponds to the NW branch of the main galaxy radial velocity distribution. The observed radial velocity gradient of the central kpc region is quite high, 135\,km\,s$^{-1}$\,kpc$^{-1}$.

\medskip
\textbf{VV\,410}
\smallskip

The radial velocity distributions have been obtained for both objects along their main axis. The satellite galaxy has a rigid body velocity distribution along almost the whole main body. Considering that the smaller galaxy is nearly edge-on, it is not possible to assure if the radial velocity curve shape obeys a matter distribution dominated by a spheroidal component, or if it is a consequence of dust opacity (see D\'iaz et al. 2001 for a similar case). The main galaxy shows a velocity distribution, along the bar orientation, with a velocity amplitude and aspect typical of spiral galaxies of this morphology type. The most perturbed side of the curve is towards the satellite galaxy.

\medskip
\textbf{VV\,350}
\smallskip

This system is conformed by a main galaxy of SABb type, nearly edge-on, and a satellite SBb, also edge-on, and no kinematic information was available till present. We obtained spectra for the major axis of both galaxies.
It is interesting to notice how the disk of the main galaxy (seen in optical images) is better defined in the zone opposite to the satellite, while the region nearer to the companion appears diffuse or perturbed, perhaps by effect of the interaction. Opposite to the direction to the satellite, the H$\alpha$ emission is higher than in the other direction. The velocity curve of the main galaxy has rigid-body rotation in most of the main body extension, although it reaches a peak toward NE, opposite to the satellite direction, while the curve in the satellite direction remains solid-body. The satellite velocity distribution has asymmetries, mainly toward the southeast, which is the region where the arm of the satellite begins that is in contact with the arm of the main galaxy.

\medskip
\textbf{NGC\,4188}
\smallskip

The satellite galaxy is located at the end of an almost non-distorted arm. With no previous radial velocity determination for the small object, its physical link with the major galaxy is confirmed in this work. As in the case of VV\,452, the satellite radial velocity is a continuation of the velocity distribution of the main galaxy. For the position angle which connects both objects we observed some asymmetries. The radial velocity values reach a maximum at 4" north-east (that is, toward the direction of the satellite) and the radial velocity distribution decreases afterwards.

\medskip
\textbf{AM\,1304-333}
\smallskip

It is clearly seen in the DSS images that the nuclei of the main galaxy is apparently displaced with respect to the geometrical center of the main galaxy disk. In 2010 we obtained broad band images with the Swope 1\,m telescope of Las Campanas Observatory. These shows that the central region is conformed by a nest of bright knots. This system seems to be an extreme case, in mass-ratio, morphology and kinematic perturbations. The blue DSS images depicts a very perturbed disk with plumes.
PA\,40$^{\circ}$: The amplitude of the radial velocity curve is 172 \,km\,s$^{-1}$. The 120\,km\,s$^{-1}$\,kpc$^{-1}$ velocity gradient is quite pronounced. The continuum emission peak is coincident with the ``turn-over" and the off-centering is 2". In the north-east direction the curve remains flat, except at 10" from the center, where a depression of about 60\,km\,s$^{-1}$ is observed. This depression could be associated with the start of the spiral arm at the tip of the bar. To the SW and after the turn-off, the radial velocity curve decreases between 8" and 15" (dust presence associated with arms?).

PA along both components: The main galaxy has a rotation curve that is notably distorted. The velocity amplitude of the main galaxy is of 75\,km\,s$^{-1}$, without inclination correction.

\medskip
\textbf{AM\,1325-274}
\smallskip

The main galaxy seems to be quite face on. The velocity distribution at PA\,166$^{\circ}$ has a steady decrease with oscillations of 40 km\,s$^{-1}$. The velocity of the satellite (which had no previous determinations) seems to be a continuation of the main galaxy rotation curve.

\medskip
\textbf{AM\,1416-262}
\smallskip

The main galaxy of this system is classified as a Seyfert\,1. Two position angles have been observed for the main galaxy and no kinematic disturbances are observed.

\medskip
\textbf{AM\,1427-432}
\smallskip

The previously unknown radial velocities of the smaller galaxy seem to be a continuation of the radial velocity distribution of the main galaxy. The rotation curve of the main galaxy has a clear asymmetry  with respect to the position corresponding to the maximum emission in the continuum, resulting in the velocity gradient being higher in the direction to the satellite.

\medskip
\textbf{VV\,452}
\smallskip

The small satellite galaxy is placed at the end of the tidal arm, with the other arm of the main galaxy being more diffuse and shorter. In this case, we also determined that both components constitute a physical pair, and also the radial velocity of the smaller galaxy results to have continuity with the radial velocity curve.
PA\,160$^{\circ}$: Although no major distortions are seen, it is clear that there are more irregularities in the side near the satellite. The observed radial velocity gradient is of 30\,km\,s$^{-1}$\,kpc$^{-1}$.
PA\,0$^{\circ}$: The location of the maximum emission in H$\alpha$ is not coincident with the peak emission in the continuum. The observed radial velocity gradient, in the radius range 0-3\,kpc, on direction to the satellite, is of 55\,km\,s$^{-1}$\,kpc$^{-1}$, being much more pronounced than in the opposite direction, with a gradient of 15\,km\,s$^{-1}$\,kpc$^{-1}$. The radial velocity of the satellite is similar to that of the nearest measured point in the main galaxy. The gradient in the rigid-body zone is higher in the satellite galaxy side. In both observed position angles, the  kinematic center does not coincide with the photometric center, being the last one determined by the peak of the continuum emission in the spectrum.

\medskip
\textbf{AM\,1955-170}
\smallskip

No previous radial velocity determinations are known for this system. We performed observations across PA\,90$^{\circ}$ for the larger galaxy, so it was not possible to confirm if this is a physical pair, although the morphology suggests it.
Unfortunately the H$\alpha$ emission is strongly contaminated with sky lines so it was not possible to obtain a radial velocity distribution although the emission in one of the arms is intense enough to derive the velocity of the main galaxy, being of (17180\,$\pm$\,30) km\,s$^{-1}$.

\medskip
\textbf{AM\,2058-381}
\smallskip

PA\,=\,9\,$^{\circ}$ (direction along the bar in the main galaxy): From the center (continuum emission peak) and towards the south-west, the curve rises as solid-body like and reaches a relative radial velocity of 195\,km\,s$^{-1}$. In the NE radius range 0-4", the curve has a gradient notably shallower than in the south-west side, decreasing only 20\,km\,$^{-1}$ with respect to the value corresponding to the central region. After 4", an abrupt jump of 115\,km\,$^{-1}$ is observed, after that, the curve remains nearly constant till 8" and has another jump of 60\,km\,s$^{-1}$. These jumps could be due to absorption effects (Zasov $\&$ Kopherskov 2003) although the model considered by these authors to explain the jumps is applied in edge-on galaxies. This effect could also be produced by warps. The branch in the curve where these steps are seen is located in the side where the tidal arm connecting to the satellite starts. The kinematic axis coincides with the direction of the bar, therefore the observed rotation curve could also be affected by the gas flows along the bar.

\medskip
\textbf{AM\,2105-332}
\smallskip

For the main galaxy it is only possible to detect H$\alpha$ emission for the central 5\,kpc. The amplitude of the rotation curve is 360\,km\,s$^{-1}$. The extension from which we can derive information is too small compared with the extension of the galaxy. Apparently, there is solid-body rotation in the central 2\,kpc. The observed radial velocity gradient in this zone is large, of 144\,km\,s$^{-1}$\,kpc$^{-1}$. No major perturbations are observed in this distribution, although beyond 2\,kpc we noted certain asymmetry. To the south-east the curve remains almost flat, while in the direction to the satellite, the curve rises 74\,km\,s$^{-1}$.

\medskip
\textbf{AM\,2256-304}
\smallskip

This system could be an extreme case of a high velocity encounter. The systemic radial velocity difference between both galaxies is 1400\,km\,s$^{-1}$. The velocity distribution for the main galaxy is symmetric, presenting oscillations in the flat region of the curve, of 50\,km\,s$^{-1}$ in the north and reaching 8\,km\,s$^{-1}$ in the southern region, where originates the tidal arm that reaches the apparent position of the satellite. This non-circular motions are probably consequence of the presence of spiral arms. The observed velocity gradient in the solid-body region is of 50\,km\,s$^{-1}$\,kpc$^{-1}$. In order to be a bound system, the center of mass of both components has to be several 10$^{12}$ \textit{M$_{\odot}$} which matches the approximate total
mass of a spiral galaxy dominated by a dark halo. The main galaxy has a diameter of $\sim$\,20\,kpc therefore a high velocity encounter is a possible scenario.

\subsection{Comparison between kinematic and photometric major axis}

In Table\,3 we list the values corresponding to the position angles of the kinematic major axis of 12 main galaxies of the sample, which were determined from the radial velocity distributions corresponding to two or more position angles. In the same table we list the photometric major axis obtained from the LEDA (these major axis were obtained considering the 25 magnitude isophote in B-band). The LEDA determinations for the direction of the major axis assign the same weight to the disk and to other structures as the tidal arm or bridges, so the photometric major axis could be ill defined in several cases. For this reason, we measured the photometric major axis in the R-band DSS images, since this band is representative of the more evolved stellar population of the disk. Table\,3 includes the position angle of the bars and the absolute values of the differences between the position angles of the major kinematic axis and the major photometric axis. Table\,3 also includes the position angle difference between the bar and the kinematic major axis. No significant discrepancies have been found, with the exception of the larger galaxy of AM\,0430-285. However the value of PA for the major axis is not far from the PA of a small bar-like subsystem of this galaxy. The bar is probably causing the difference between the photometric and kinematic position angle, since the bar could produce non-circular motions. In general, it was found that the PA of the kinematic and photometric major axis are similar and if an important deviation is measured, the galaxy has a bar. In conclusion, there is no evidence that in this type of object the interaction is strongly affecting the alignment of the photometric and kinematic major axis, as would be the case in severely warped disk systems.

\subsection{Satellite orbital plane}
In Table\,4 we detail the position angle of the photometric major axis of the main galaxies, the position angle of the center of the satellites, as measured from the centers of the main galaxies, as well as the difference between these values. In Figure\,5 we show a histogram of the PA difference. About half of the sample presents PA difference smaller than 20$^{\circ}$, so in a statistically sense, this could suggest that satellite galaxies are closer to the main galaxy plane. For verification, we have applied a Monte Carlo method, by running a set of simulations for random position angle values, between 0 and 90$^{\circ}$. Each simulation generates 21 random values in PA difference. For each set we considered its average value, which for the real case is about 28$^{\circ}$. After running 1500 random sets, we obtained only 3 average PA differences under 30$^{\circ}$: 25.5$^{\circ}$, 28.6$^{\circ}$ and 29.5$^{\circ}$. This means that the probability of the distribution being random is less than 0.002, and the satellite galaxy orbits are significantly closer to the galaxy equatorial plane during the M51 phase of the interaction.

\subsection{Mass determinations}

 We determined masses in Keplerian approximation for several members of the present catalogue, mainly for the larger components and also for some satellite galaxies. The estimated masses have been obtained for the outermost spectroscopic measurement and considering the maximum amplitude for each velocity distribution. We performed the corresponding correction for inclination of the disk and deviation from the observed position angle with respect to the direction of the major axis. The results are listed in Table\,5. At the top of Figure\,6 we present the distribution of the main galaxy masses. Ks-band M/L$_{K}$ ratios were calculated using the Keplerian dynamical masses derived from the kinematics, and the integrated Ks-band luminosities were determined by aperture photometry at the 2MASS Ks-band images. \\
For comparison, we considered the galaxies included in the catalogue of Karachentsev et al. (2004). This sample consists of 451 Local Volume galaxies (these authors considered objects with distance estimates of \textit{D} $\lesssim$ 10 Mpc or radial velocities \textit{V}$_{LG}$ $<$ 550 km\,s$^{-1}$). Karachentsev \& Kutkin (2005) obtained Ks-band luminosities for the mentioned galaxies from 2MASS and studied their relation between Mass/Ks-band Luminosity vs. Ks-band luminosity. As we show in Figure\,6, the range of M/L$_{K}$ ratios is in very good agreement with the range of values determined by Karachentsev \& Kutkin (2005), who derive the luminosities from 2MASS but calculate the dynamical masses from HI line widths. However, the M51 sample is located in the same region of the M/L$_{K}$ vs. L$_{K}$ plot in which are located the brightest galaxies from Karachentsev et al.'s sample. These authors show that the brightest nearby galaxies detach from the normal distribution of objects, in a sub-sample with log L$_{K}$ $\sim$ 10.5, comparable with the M51-type galaxy luminosity about log L$_{K}$ $\sim$ 10.8.\\
We calculated the M/L$_{K}$ average for the M51-type galaxy sample, M/L$_{K}$ $\sim$ -0.08, rather below of that corresponding to the Karatchensev \& Kutkin (2005)'s linear regression at log L$_{K}$ $\sim$ 10.8 (M/L$_{K}$ $\sim$ 0.21). In the plot, it can be seen that four galaxies have a distinctive low M/L$_{K}$ ratio. We look for evidence of exalted star formation or kinematic distortion as a possible explanation for the low M/L$_{K}$ observed.
These objects with very low M/L$_{K}$ are identified as the main galaxies of AM\,1304-333, whose morphology is very distorted and has a spectrum indicative of enhanced star-forming activity (van den Broek et al 1991; Gunthardt 2009), NGC\,633, already classified as a nuclear starburst galaxy (Gunthardt 2009), AM\,0430-285, main galaxy which has not available spectrum data in the literature and the satellite galaxy of NGC\,633, whose nuclear spectrum is characterized by an enhanced star-forming activity (Gunthardt 2009). The mean M/L$_{K}$ ratio of the M51-type galaxies without these objects is $\sim$ 0.15, very close of the expected value from the Karachentsev \& Kutkin (2005) sample.

\subsection{Tully-Fisher relation}
We studied the Tully-Fisher (T-F) relation for some of the objects of the present sample. We corrected the maximum velocities for inclination and selected B magnitudes from the LEDA, which are corrected for galactic and internal extinction. The T-F relation for M51-type galaxies in B-band may be fitted by M(B)\,=\,((-2.69\,$\pm$\,0.78)\,$\cdot$\,(logV$_{max}$)) - (14.86\,$\pm$\,1.60). In Figure\,7, we compared our fitting for M51-type galaxies with the T-F relation obtained by Kudrya \& Karachentseva (2012) for 169 edge-on isolated galaxies (these objects are listed in the 2MIG catalogue of isolated galaxies; Karachentseva et al. 2010) resulting. The T-F fitting for M51-type galaxies is flatter than the relation corresponding to isolated galaxies. As can be observed in Figure\,7, upper left, the slope corresponding to the fitting for member galaxies of M51-type systems is also flatter than the slope of the classic T-F relation (Pierce $\&$ Tully (1992); Tully et al. (1998); Tully $\&$ Pierce (2000)). Reshetnikov $\&$ Klimanov (2003) (RK) also obtain, for a sample of M51-type galaxies, a T-F flatter than the relation corresponding to nearby galaxies. If we add the data from RK to our own data in the T-F relation, the same trend is observed and may be fitted by M(B)\,=\,((-3.20\,$\pm$\,0.64)\,$\cdot$\,(logV$_{max}$)) - (13.56\,$\pm$\,1.11). Comparing our sample with the data of RK, only three objects are in common, the main and satellite galaxies of VV\,350 and the main galaxy of VV\,452. In Figure\,7, upper right, the galaxies of both samples are included, although considering that the only source of the RK data is their Figure\,4, that is, we cannot discriminate to which object corresponds each TF point in the plot of RK, we decided not to include the data corresponding to VV\,350 and VV\,452 of our sample. The fitting of our sample is flatter than the fitting of RK, due to the sample of RK including two objects of lower masses and luminosities, making their fit steeper than ours. As may be seen, the brightest galaxies are placed near the standard relation, which is compatible with the fact that for the most massive systems, the T-F relation does not depend on the environment (Barton et al. 2001). The less massive galaxies, members of M51-type systems, have a higher luminosity than would correspond to a standard T-F relation. The average B-band flux ratio is of $\sim$ 4, between the M51-type galaxies luminosity with velocities lower than log(V$_{max}$)\,=\,2.1), and the luminosity expected from the T-F fitting of the isolated galaxies of the Kudrya \& Karachentseva (2012) sample.    \\
Following this line of thought, we add a study of the T-F relation in the Ks-band, since this band may be less affected by recent star formation. The result (bottom of Figure\,7) is also a flatter slope when compared with the fitting of 182 edge-on isolated galaxies of Kudrya \& Karachentseva (2012). For comparison, the T-F relation found by Torres-Flores et al. (2011; GHASP spiral galaxies) are also added in Figure\,7, as well as the fitting of Masters, Springob $\&$ Huchra (2008) for 2MASS galaxies.  \\
The average Ks-band flux ratio is of $\sim$ 6 for M51-type galaxies with radial velocities lower than log(V$_{max}$)\,=\,2.1), with respect to the T-F fitting in Ks-band for the sample of Kudrya \& Karachentseva (2012).\\
A similar situation is observed in the study of members of Hickson compact groups of galaxies (Torres-Flores et al. 2013), where a few low mass outlier objects deviate from the standard T-F relation for field galaxies. We noticed that these outliers are occupying the same region in the plot of the T-F relation as the low-mass M51-type galaxies. This happens for both B-band and Ks-band T-F relations.
According to Torres-Flores et al. (2013), the deviations from the standard relation could be due to a combination of a flux increment generated by the enhancement of star-forming events with the uncertainties in the determination of the kinematic widths due to kinematic distortions induced by gravitational interactions. Torres-Flores et al. (2013) observe for some outliers a larger deviation in Ks-band than in B-band. They notice that, according to Maraston (1998), if thermally pulsating asymptotic giant branch stars (TP-AGB) are considered in stellar population synthesis models, they may make an important contribution in the K band. In particular, Maraston (1998) found that this effect is significant for an evolved burst with an age of about 700 Myrs. \\
 Another study of objects that are suffering gravitational interactions is that of Barton et al. (2001), who determine the T-F relation for galaxy pairs. They also observe for low-mass objects an excess in the B-band flux, with respect to the one expected for standard T-F relations. These authors determined that in extreme cases the deviation from the standard T-F relation may be explained by uncertainties in inclination, in whose cases long tidal tails should be observed. Since M51-type galaxies in general present long tidal tails, in our sample the effects due to indetermination in inclination could be larger than those of Barton et al.'s sample. Barton et al. (2001) also proposed that another possible explanation for the outliers is dissipative effects on the rotation curve (gas infall and non-uniform or truncated gas emission).\\
 Our M51-type galaxy sample includes low-mass objects with asymmetric curves (e.g.: AM\,0459-340, AM\,1304-333, NGC\,633), truncated rotation curves (NGC\,633 satellite, NGC\,4188), and star-forming enhancement (AM\,0459-340, main and satellite galaxies of NGC\,633, AM\,1304-333). In addition, we must take into account that in objects with well-marked signs of gravitational interactions the determination of the major axis is more tricky, possibly resulting in a underestimation of the rotation curve amplitude. A next step on this subject could be to perform a study using 3-D spectroscopy techniques. There are few low-mass M51 objects studied, so it would also be desirable to add observations of less massive systems.\\

\subsection{Synthetic radial velocity distribution}\label{synt}
For future numerical simulations of the gravitational encounter between main and satellite galaxies (components of a typical M51-type system), it is necessary to consider a representative radial velocity curve of the larger component. One possible approach is to derive a synthetic curve from the available radial velocity data for a set of main galaxies. For this purpose, we selected from our catalogue the radial distributions along major axis position angle for 13 systems (data\_{a}, table\,6), retaining the systems which present less ambiguity in the determination of the major axis position angle and the inclination. The radial velocity data have been deprojected, considering the inclination of the main galaxy. Being more restrictive, we excluded some systems and built two other data sets, data\_{b} (11 objects) and data\_{c} (9 objects), being the last set of data the one with more constrained parameters, so this last set of radial velocities contains the most ``trusty" data values. The data have been normalized, in both radial velocity and position, dividing by the peak value in radial velocity, and in the case of the position, dividing by the disk  extension of the main component in R-DSS band.\\
For fitting the data, we have used the software ``TableCurve". The functions\footnote{A description of the functions used in this fitting process, as well as their coefficients (`a', `b', `c', `d', `e', `f'), may be found in http://systat.co.kr/products/TableCurve2D/help/1041.html.} that resulted in a best fitting for each set of data are: `Asymmetric Double Gaussian Cumulative' (ADC) for data\_{a}, `Modified Gaussian' (GaussMod) for data\_{b} and `Symmetric Double Gaussian Cumulative' (SDC) for data\_{c}. In Table\,7 we list the coefficients of the functions that fitted the data for each set (Figure\,8). In the same table we include the standard deviation ($\sigma$). Also, for each data set, and as a simple visual test of the fitting, we divided the range of normalized radial distance in different bins, obtaining an average value of normalized radial velocity for each bin. The functions obtained with the mentioned software provided a good fitting to the average values. The errors associated to the average values correspond to the standard deviation and are plotted in Figure 8. As can be seen in Figure\,8d, there is not significant difference between the fittings of the three sets of data. \\

As a different approach, we have built a representative gravitational potential of the spiral galaxies corresponding to set ``c". This spiral galaxy model consists of a disk described by an exponential law (Freeman (1970)), plus a dark halo represented by a NFW mass distribution (Navarro, Frenk $\&$ White 1996). The choice of these potentials corresponds to the description of Barnes $\&$ Hibbard (2009) for the numerical modeling of interacting disk galaxies. Considering that we have obtained radial velocities from long-slit spectroscopy (with half arcsecond uncertainties in the galaxy center position), the galactic centers are affected by important interstellar extinction (underestimated radial velocities), and considering the seeing of the observations, we conclude that the central radial velocities have significant uncertainties. Therefore, the resolution obtained for small radii is not adequate to properly trace the mass of the bulge component. Consequently, we only have considered the disk and halo components from the dynamic model proposed by Barnes $\&$ Hibbard (2009). In order to resolve the well-known disk-halo conspiracy in the rotation curve description, we have assumed the parameter relation proposed by Barnes $\&$ Hibbard (2009) for both components. The best description for the circular velocities, according to the adopted model, is shown in Figure\,9a.
Assuming for the set ``c" sample a maximum circular velocity and maximum radius average values of 190 km s$^{-1}$ and 11.75 km s$^{-1}$ respectively, the total mass of the disk is M$_{disk}$\,=\,(4.75\,$\pm$\,0.35)\,x\,10$^{10}$\,M$_{\odot}$ with a scale radius of \textit{a}$_{d}$\,=\,(3.5\,$\pm$\,0.2)\,kpc, while the halo mass inside the scale radius is of M$_{halo}$(\textit{a}$_{h}$)\,=\,(4.0\,$\pm$\,1.1)\,x\,10$^{10}$\,M$_{\odot}$ with scale radius of \textit{a}$_{h}$\,=\,(10.5\,$\pm$\,0.3)\,kpc. We listed in Table\,7 the last parameters, as well as the standard deviation of fitting. In order to take into account the observational uncertainties in the central region, we have performed a second fitting of the data just beyond R=1.5 Kpc, including a spherical potential (Hernquist 1990), as proposed by Barnes $\&$ Hibbard (2009), which contains 5$\%$ of the total mass and has a scale radius of 24$\%$ respect to that of the disk component (Barnes $\&$ Hibbard 2009). In this case, the fitting parameters were able to differ from the Barnes $\&$ Hibbard (2009) relations in less than 25$\%$ in mass and less than 10$\%$ in scale radius. The mass model parameters including a bulge component are the following: M$_{bulge}$\,=\,1.12\,x\,10$^{10}$\,M$_{\odot}$, \textit{a}$_{b}$\,=\,0.85\,kpc; M$_{disk}$\,=\,3.5\,x\,10$^{10}$\,M$_{\odot}$, \textit{a}$_{d}$\,=\,3.5\,kpc; M$_{halo}$(\textit{a}$_{h}$)\,=\,3.8\,x\,10$^{10}$\,M$_{\odot}$, \textit{a}$_{h}$\,=\,9.5\,kpc. Thereby, a comparison between this synthetic rotation curve and higher resolution rotation curves is possible. This synthetic rotation curve serves as a model starting point for numerical simulations of M51-type galaxies.

In order to further take into account the dispersion of observational data, we have determined two rotation curve envelopes which enclosed the individual rotation curves of set ``c" (Figure\,9c). The envelope curves keep the same scale parameters but differ in mass by 50\,$\%$.
Another useful description for these rotation velocities is a simple linear function fitting which could be easily compared with other samples. The linear fitting consists of two straight lines, that is, one that represents the central rigid rotation (R\,$<$\,2\,kpc) and another which represents the flat region of the rotation curve (R\,$>$\,3\,kpc). In Figure\,9d we show the fitting of both linear functions. The central region has a radial velocity gradient of 68\,km\,s$^{-1}$\,kpc$^{-1}$ while the flat region shows a slight increase with radius of 1.1\,km\,s$^{-1}$\,kpc$^{-1}$ (this slight slope is consistent with the previous synthetic description that includes a large scale halo component).

\section{Conclusions}
From the present kinematic sample consisting of 21 interacting M51-type galaxy systems, we found that the radial velocity distributions present asymmetries in about half of the studied main galaxies. Typical examples are AM\,0459-340, VV\,410 or VV\,452.
In some systems a detailed scrutiny reveals subtle asymmetries: at first sight the curve of AM\,0639-582 is normal, however, the observed radial velocity gradients are different in both sides of the distribution, taking as reference the position corresponding to the red continuum emission peak in the spectra.\\
Only AM\,1304-333 has a very distorted radial velocity curve. We also observe an important morphological perturbation, which would be the result of an ongoing multiple merger. \\
 For almost all the main galaxies that belong to the present sample, the observed perturbations are not as significant as the ones involved in fusions. In fact we found only small or moderate differences between the photometric and kinematic major axes of the main galaxies.
 In systems with low ratio between satellite and main galaxy masses, the systematic velocities of the satellites are a continuation of the velocity curve of the larger galaxy, which agrees with the results obtained by Reshetnikov $\&$ Klimanov (2003): The relative velocity of the satellite is approximately equal to the velocity in the outer parts of the main galaxy disk.\\
 M51-type galaxies have average luminosities (log L$_{K}$ $ \sim$ 10.5) that are comparable to the luminosities of the brightest objects in the nearby galaxy sample (451 objects) of Karachentsev et al. (2005). This may be due to a selection bias (we are detecting only the brightest M51 galaxies), but the importance of this effect seems diminished by the fact that only the prototype M51 (NGC\,5194$/$95) in Karachentsev$'$s sample match the type of objects considered here, and that this prototype system is also a luminous one. The galaxy brightness in the K-band could be raised by an enhancement of the young stellar population in these galaxies, shifting the location of M51 galaxies towards lowered M/L$_{K}$ ratios in the plot. This effect is seen for some less massive objects in the Ks-band Tully-Fisher relation. Therefore the M51 phenomenon happens in intrinsically massive bright spiral galaxies (as the case of M51 itself), which notwithstanding have quite normal M/L$_{K}$ ratios.\\
 The B-band and K-band Tully-Fisher relation for these systems are flatter than the standard one derived for isolated galaxies. The same trend is observed in other interacting galaxy samples. More data of low mass objects and 3D spectroscopic studies would be desirable for improving the T-F characterization of this type of objects.\\
From the kinematic data, we have derived a synthetic radial velocity distribution that serves as reference for future numerical simulations. The synthetic rotation curve is near to solid body-like inside 4 kpc, and then is nearly flat within the radial range 5-15 kpc. The semi-amplitude of the curve is $\sim$ 190 km\,s$^{-1}$\,kpc$^{-1}$. \\
The distribution of the position angle difference between the main galaxies major axis and the segment that connects the centers of main and satellite galaxies, shows an excess of companions located towards the main galaxy major axis. A selection effect of M51 companions located nearer to the galaxy major axis can be discarded as there is no correlation with the system inclination. This indicates that in most of the systems in the M51 phase of the interaction, the companion orbit has a large projection on the main galaxy equatorial plane. Consistently, the radial velocity differences between main and satellite galaxies, normalized to the peak radial velocity of the velocity distribution of the main galaxy, indicates that the orbital motion of the satellite is within the range of amplitudes of the main galaxy rotation curve. All the studied M51 systems are therefore gravitationally bound, except for one pair which seems to be a high velocity encounter.

Despite the significant increase in data that resulted from the present work, there is a clear need of spectroscopic observations with higher spatial resolution and of a larger sample, for better constraining the kinematics and orbital parameters of M51-type galaxies.\\

\section*{Acknowledgements}
It is a pleasure to thank the referee whose suggestions helped to improve the contents and presentation of this paper.
This work is based on observations obtained at Complejo Astron\'omico El Leoncito (San Juan).
GG wants to specially thank Dr. Estela L. Ag\"uero for her suggestion of studying the M51-type systems and for her supervision at the beginning of this research. This research has made use of the NASA/IPAC Extragalactic Database (NED) which is operated by the Jet Propulsion Laboratory, California Institute of Technology, under contract with the National Aeronautics and Space Administration. This publication makes use of data products from the Two Micron All Sky Survey, which is a joint project of the University of Massachusetts and the Infrared Processing and Analysis Center/California Institute of Technology, funded by the National Aeronautics and Space Administration and the National Science Foundation.
We acknowledge grant support from Secyt, UNC (05/N030) and grant support from CONICET (PIP 0523), ANPCyT (PICT 835).

\begin{referencias}
\reference Arp, H., 1966, \apjs, 14, 123
\reference Arp, H., $\&$ Madore, B. 1987, A catalogue of Southern Peculiar Galaxies and Associations. (Cambridge: Cambridge Univ. Press)
\reference Barnes, J. E., \& Hibbard, J.E. 2009, AJ, 137, 3071
\reference Barton, E. J., Geller, M. J., Bromley, B. C., van Zee, L. \& Kenyon, S.J. 2001, AJ, 121, 625
\reference Bushouse, H. 1987, AJ, 320, 49
\reference Calzetti, D., et al. 2005, ApJ, 633, 871
\reference D\'iaz, R. J., Rodrigues, I., Dottori, H. \& Carranza, G. 2001, AJ, 119, 111
\reference Dopita et al. 2002, ApJS, 143, 47
\reference Ellison, S. L., Patton, D. R., Simard, L. \& McConnachie, A. W. 2008, AJ, 135, 1877
\reference Fuentes-Carrera, I., Rosado, M., Amram, P., \& Laurikainen, E. 2007, A\&A, 466, 847
\reference Freeman, K. C. 1970, ApJ, 160, 811

\reference G\"unthardt, G. I. 2009, Ph.D. Thesis., Famaf, Universidad Nacional de C\'ordoba

\reference Hernquist, L. E. 1990, ApJ, 356, 359

\reference Jokim\"aki, A., Orr, H., \& Russell, D. G. 2008, Ap\&SS, 315, 249

\reference Karachentsev, I. D., Karachentseva, V. E., Huchtmeier, W. K., \& Makarov, D.I. 2004, AJ, 127, 2031

\reference Karachentsev, I. D. \& Kutkin, A. M. 2005, Astronomy Letters, 31, 5

\reference Karachentseva, V. E., Mitronova, S. N., Mel'nik, O. V. \& Karachentsev, I. D. 2010, Astrophysical Bull., 65, 1

\reference Keel, W. C. 2004, ApJS, 106, 27

\reference Klimanov, S. \& Reshetnikov, V. 2001, A\&A, 378, 428

\reference Klimanov, S., Reshetnikov, V., \& Burenkov, A. 2002, Astronomy Letters, vol. 28, No. 9, p. 579

\reference Kudrya, Y. N. \& Karachentseva, V. E. 2012, Astrophysics, 55, 435

\reference Laurikainen, E., Salo, H. $\&$ Aparicio, A. 1998, A$\&$AS, 129, 517

\reference Maraston, C. 1998, MNRAS, 300, 872

\reference Masters, K. L., Giovanelli, R. \& Haynes, M. P. 2003, AJ, 158, 174.

\reference Masters, K. L., Springob, C. M. $\&$ Huchra, J. P. 2008, ApJ, 135, 1738

\reference Navarro, J. F., Frenk, C. S., \& White, S. D. M. 1996, ApJ, 462, 563

\reference Pierce, M. J. \& Tully, B. R. 1992, ApJ, 387, 47

\reference Poggianti, B. M. 1997, A$\&$AS, 122, 399

\reference Rampazzo, R., Plana, H., Amram, P., Bagarotto, S., Boulestex, J. \& Rosado, M. 2005, MNRAS, 356, 1177

\reference Reshetnikov, V. \& Klimanov, S. 2003, Astronomy Letters, vol. 29, 429

\reference Schlafly, E. F. \& Finkbeiner, D. P. 2011, ApJ, 737, 103

\reference Smith, B. J., Struck, C., Hancock, M., Appleton, P. M., Charmandaris, V. \& Reach, W. T. 2007, AJ, 133, 791

\reference Thomasson, M. 2004, in ASP Conf. Ser. 320, The Neutral ISM in Starburst Galaxies, ed. S. Aalto, S. H\"uttemeister, \& A. Pedlar (San Francisco, CA: ASP), 81

\reference Torres-Flores, S., Epinat, B. Amram, P., Plana, H. \& Mendes de Oliveira, C. 2011, MNRAS, 416, 1936

\reference Torres-Flores, S., Mendes de Oliveira, C., Plana, H., B. Amram, P. \& , Epinat, B. 2013, MNRAS, 432, 3085

\reference Tully, B. R., Pierce, M. J., Huang, J. S. et al. 1998, AJ, 115, 2264

\reference Tully, B. R. \& Pierce, M. J. 2000, ApJ, 533, 744

\reference van den Broek, A. C., van Driel, W., de Jong, T., Lub, J., de Grijp, M. H. K. \& Goudfrooij, P. 1991, A\&AS 91, 61

\reference Vorontsov-Velyaminov, B. A. 1975, AZh, 52, 692

\reference Woods, D. F., \& Geller, M. J. 2007, AJ, 134, 527

\reference Zasov, A. \& Kopherskov, A. 2003, Astronomy Letters, 29, 437

\end{referencias}

\section{Appendix}

In what follows we list the functions employed in $\S$3.7. The coefficients `a', `b', `c', `d', `e' and `f' are detailed in Table\,7:\\


\textbf{``Asymmetric Double Gaussian Cumulative'' (ADC)}\\


\[y = \frac{\displaystyle a}{\displaystyle 4} \left [1 + erf \left ( \frac{\displaystyle x\,-\,b + c/2}{\displaystyle d\sqrt{ 2 }} \right ) \right ] \left [\frac{\displaystyle 1}{\displaystyle 2} - \frac{\displaystyle 1}{\displaystyle 2}\,erf \left ( \frac{\displaystyle x\,-\,b - c/2}{\displaystyle \sqrt{ 2 }e} \right ) \right ] + f\]\\\\


\textbf{``Modified Gaussian'' (GaussMod)} \\


\[y = e + a\,\, exp\left [ -\,\frac{\displaystyle 1}{\displaystyle 2} \left ( \frac{\displaystyle \left|x-b\right|}{\displaystyle c} \right )^{d} \right ]\]\\

\textbf{``Symmetric Double Gaussian Cumulative'' (SDC)}\\


\[y = \frac{\displaystyle a \left [1 + erf \left ( \frac{\displaystyle x\,-\,b + c/2}{\displaystyle d\sqrt{ 2 }} \right ) \right ] \left [1 - erf \left ( \frac{\displaystyle x\,-\,b - c/2}{\displaystyle d\sqrt{ 2 }} \right ) \right ]}{\left [\displaystyle 1 + erf \left ( \frac{\displaystyle c}{\displaystyle 2d\sqrt{ 2 }} \right ) \right ]^2} + e\] \\\\


where \textit{erf(x)} is the Gauss error function,

\[erf(x) = \frac{\displaystyle 2}{\sqrt{\displaystyle \pi }}{\int_{0}^{x} \! e^{t^{2}}  \,dt} \]

\begin{figure}

  \centering
  \includegraphics[width=0.7\textwidth]{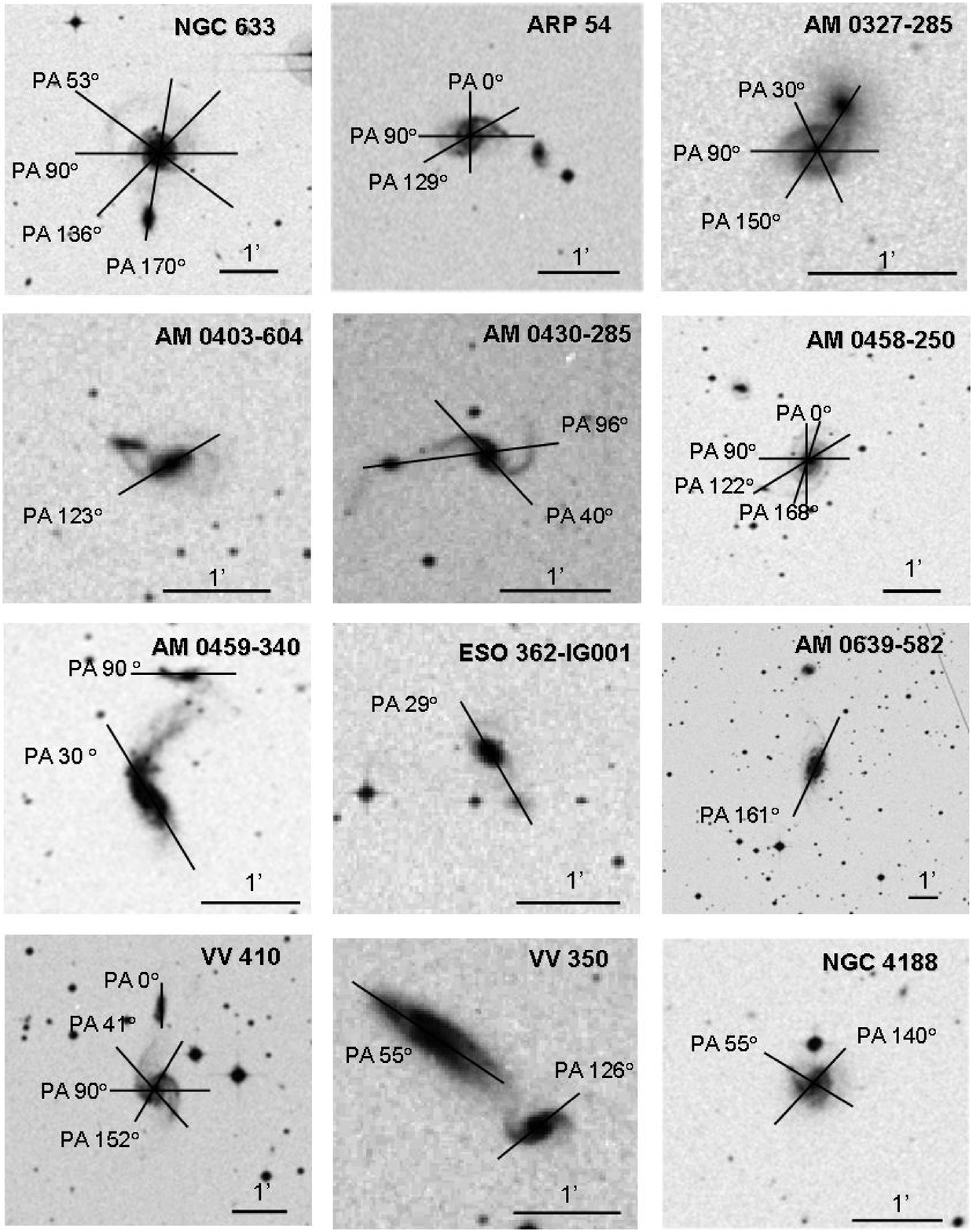}~\hfill
\clearpage
  \includegraphics[width=0.7\textwidth]{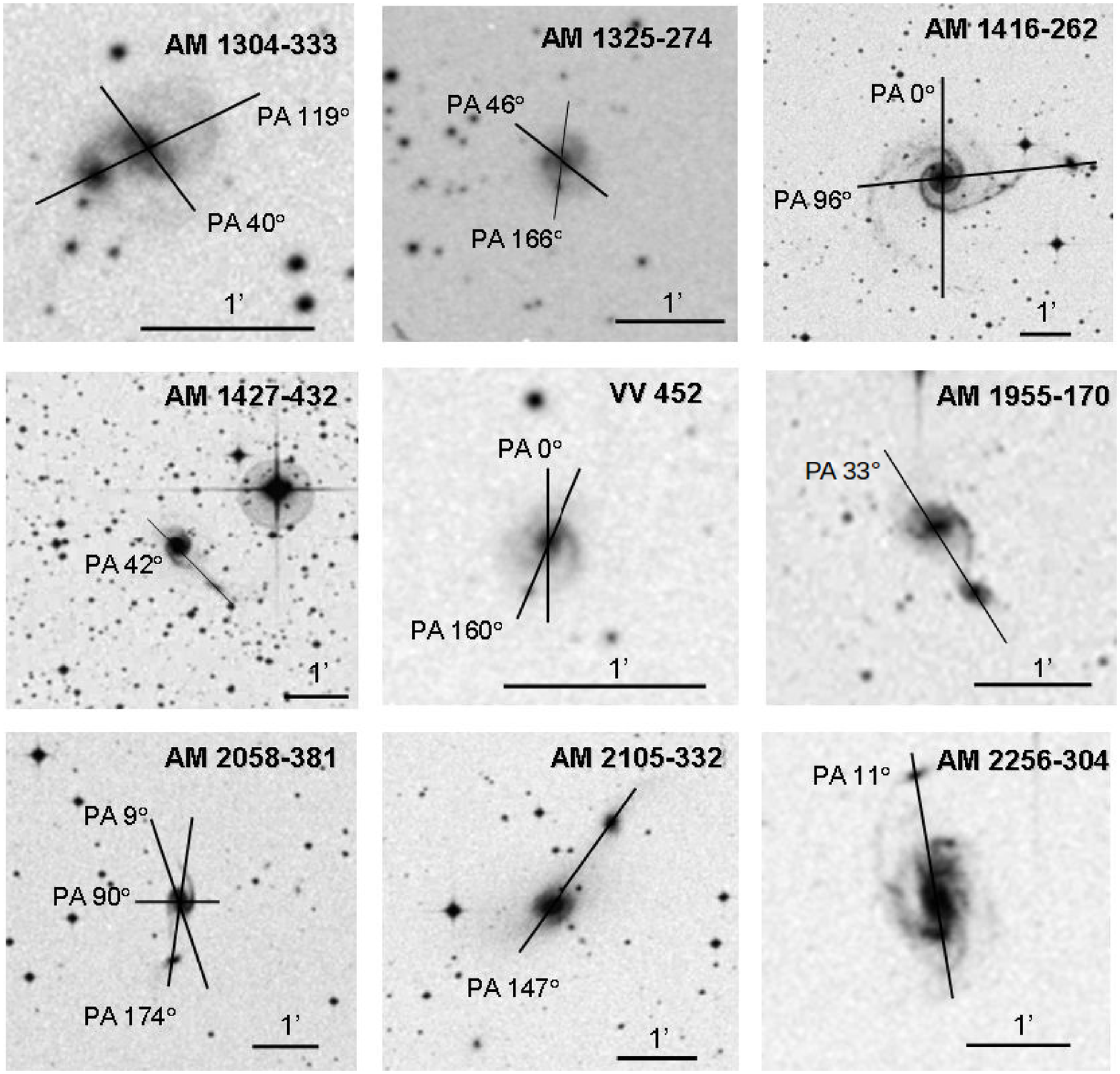}
  \caption{B-band images from the \textit{``Digitized Sky Survey"}. The dark segments mark the position angles of the obtained spectra.}

\end{figure}

\clearpage

\begin{figure}[!ht]
  \centering
  \includegraphics[width=0.45\textwidth]{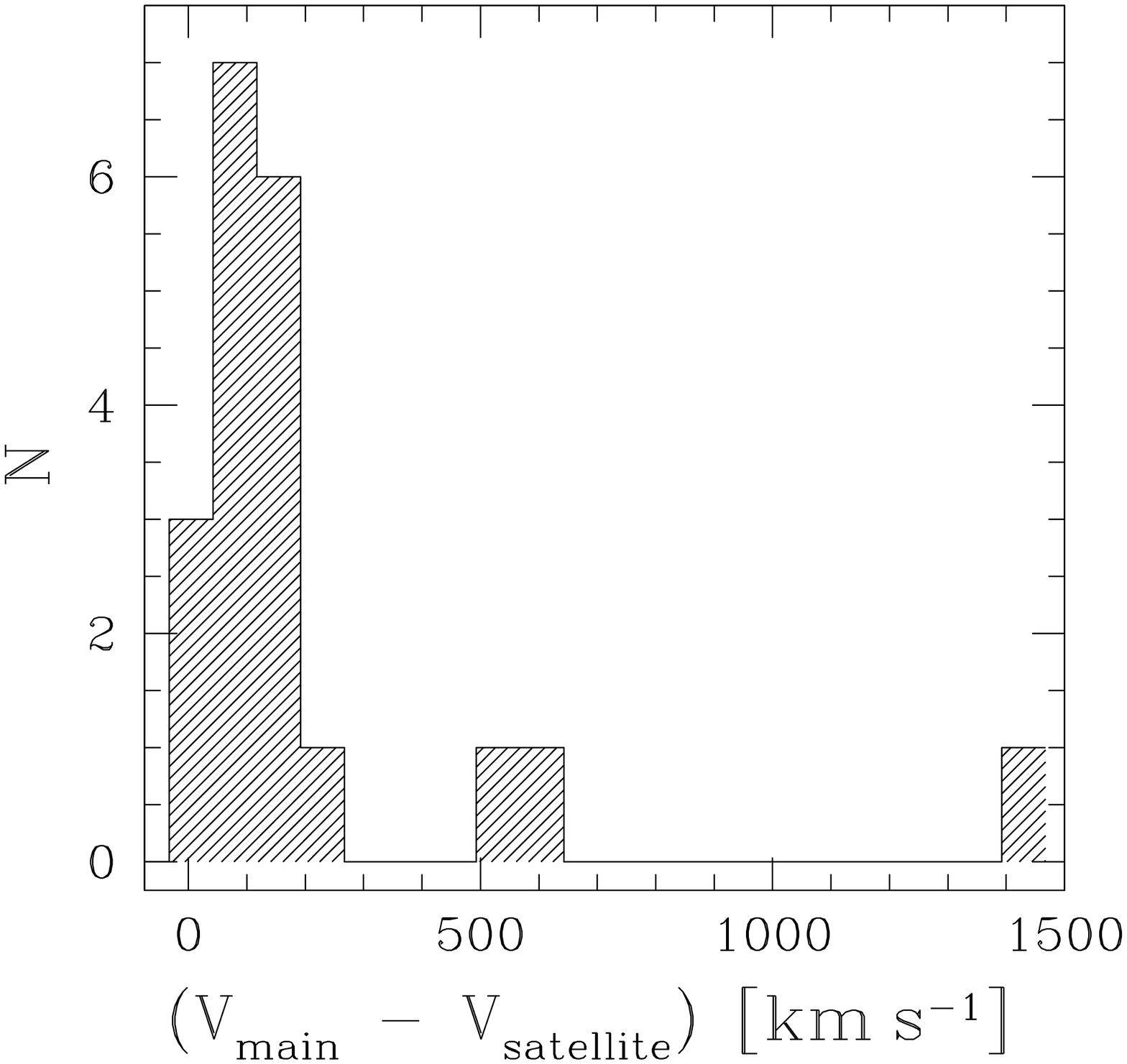}
   \includegraphics[width=0.45\textwidth]{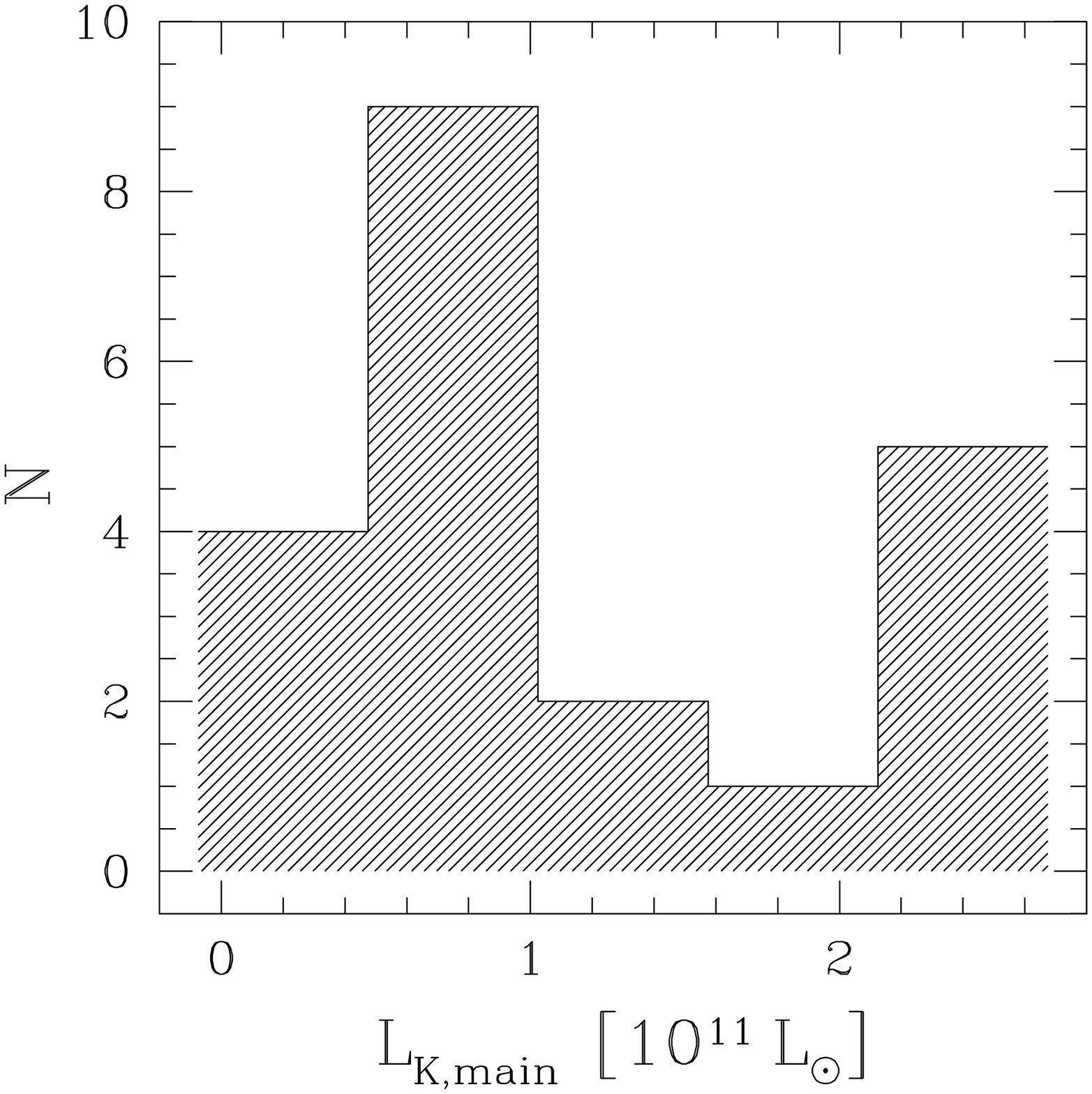}
  \includegraphics[width=0.45\textwidth]{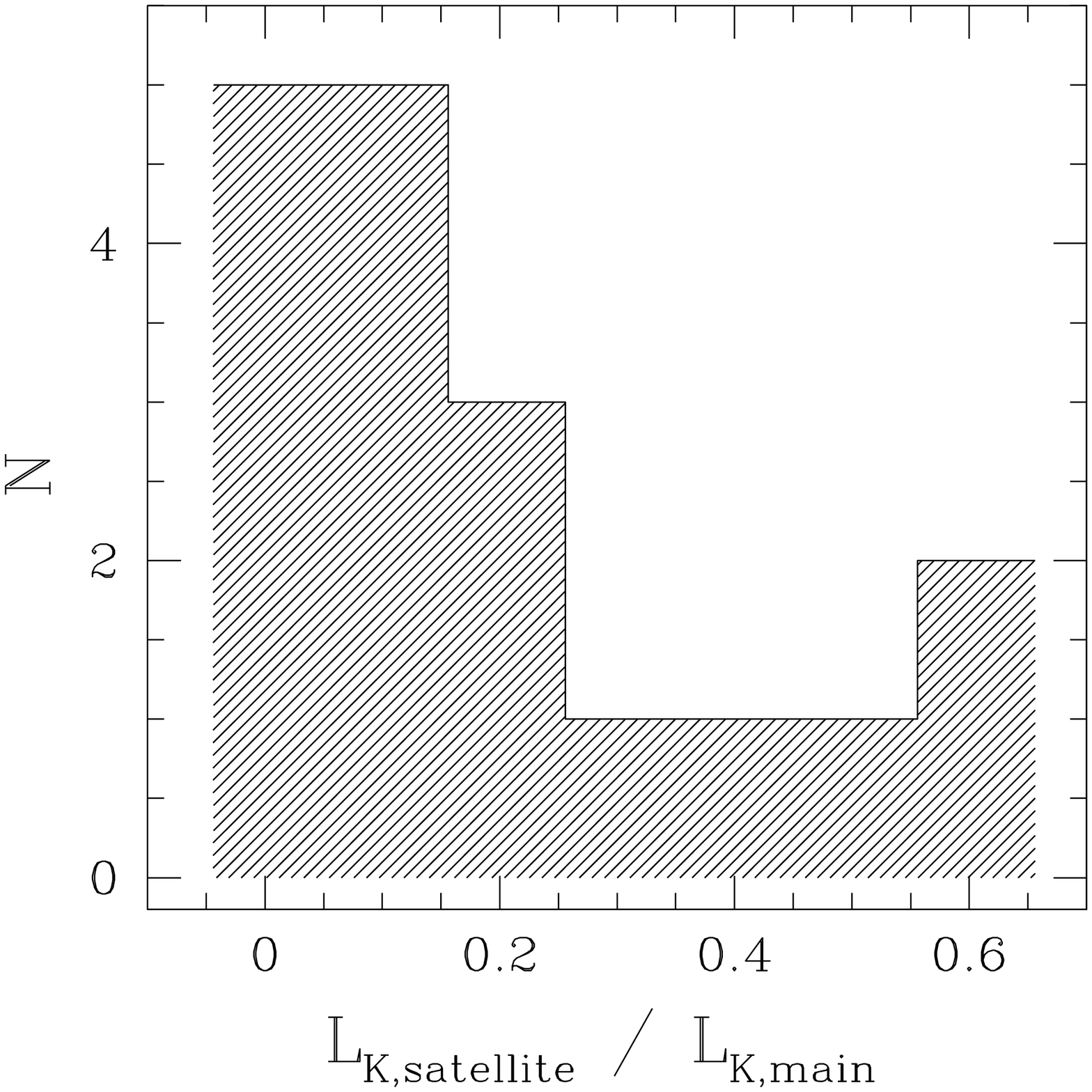}
  \caption{\textit{Upper left}: Distribution for the modulus of the radial velocity difference between main and satellite galaxies. \textit{Upper right}: Distribution of the Ks-band luminosity (2MASS) for the main galaxies of the present sample. \textit{Bottom}: Distribution of the ratio of luminosities in Ks-band (2MASS) for the satellite galaxy Ks-band luminosity with respect to the main satellite Ks-band luminosity.}

\end{figure}

\clearpage

\begin{figure}[!ht]
  \centering
  \includegraphics[width=0.8\textwidth]{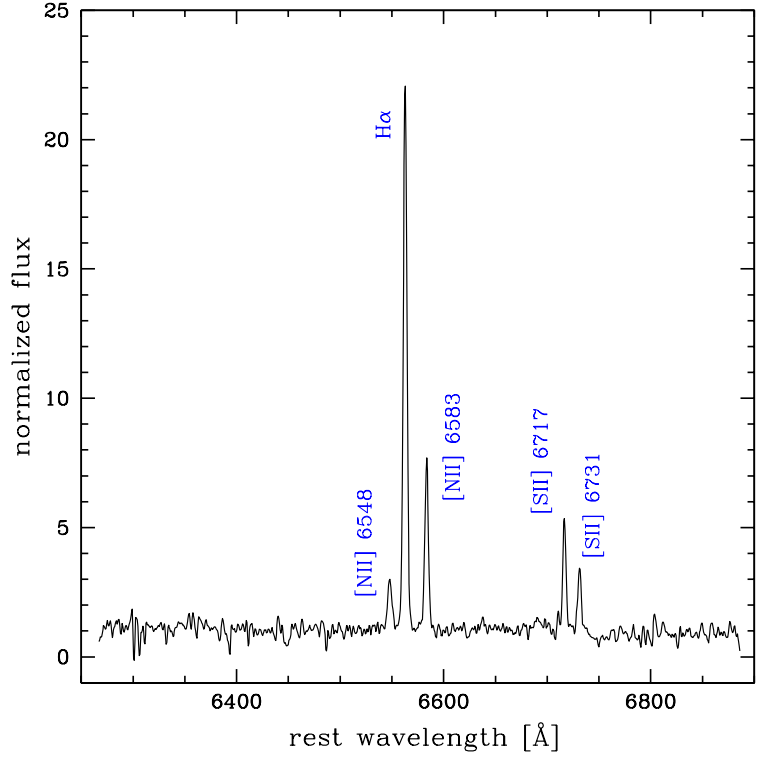}

  \caption{Example of an optical spectrum obtained with the REOSC spectrograph, in CASLEO, Argentina, with a 1200 l/mm grating. It corresponds to the nuclear region of the AM\,0450-340 main galaxy. The spectrum extraction is of 3 arcsec wide. The abscissa is in arbitrary units and is normalized to the spectrum continuum.}

\end{figure}

\clearpage

\begin{figure*}
\centering
  \includegraphics[width=17cm]{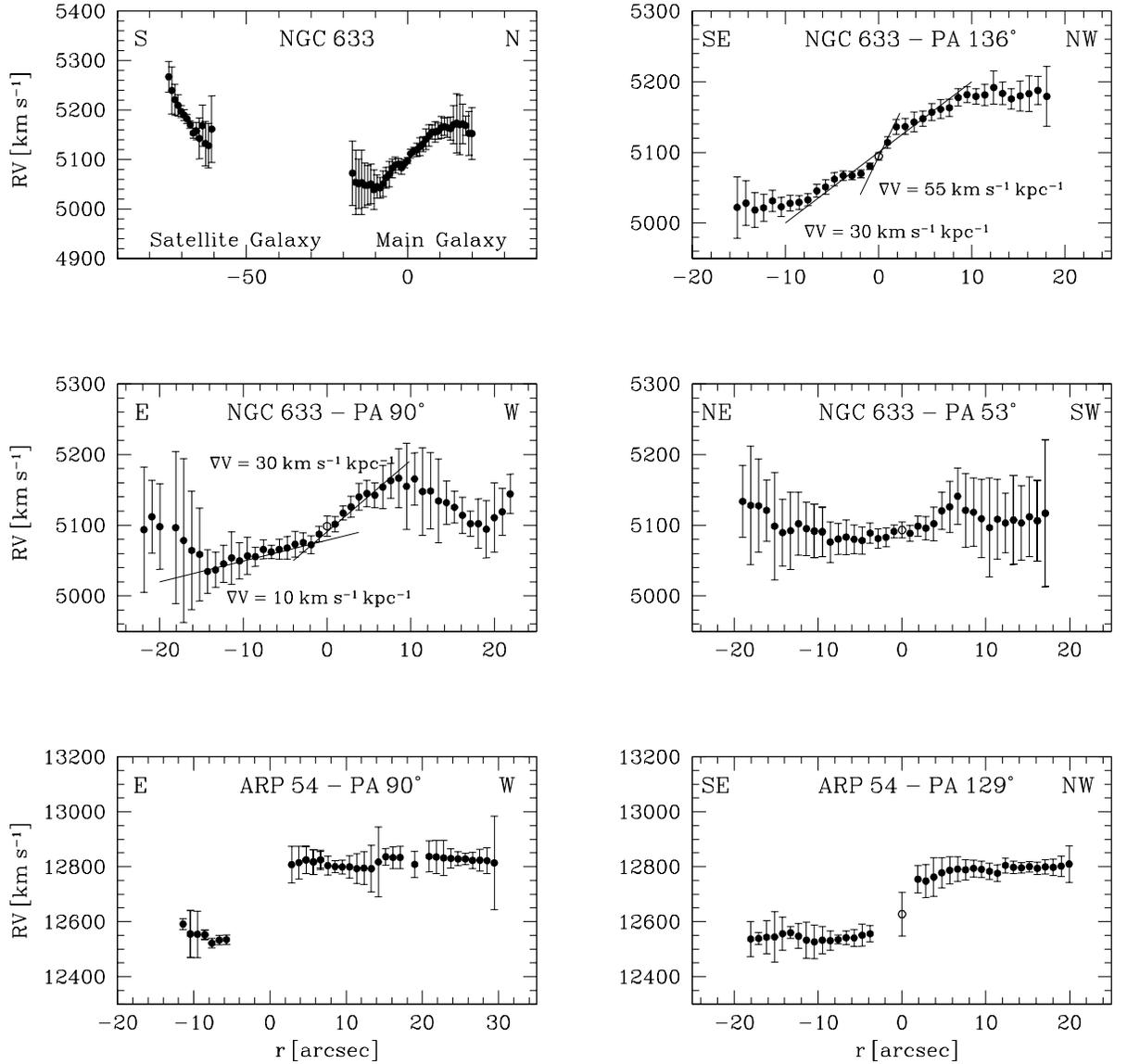}
    \caption{Heliocentric radial velocity distributions for objects of the M51-type galaxies catalogue. Open circles mark the location of the continuum emission peak in the spectra at the rest frame wavelength range 6400-6800 \AA. In some cases, the best-fit to the velocity gradient is drawn as a straight line.}

\end{figure*}

\addtocounter{figure}{-1}
\begin{figure*}
\centering
  \includegraphics[width=17cm]{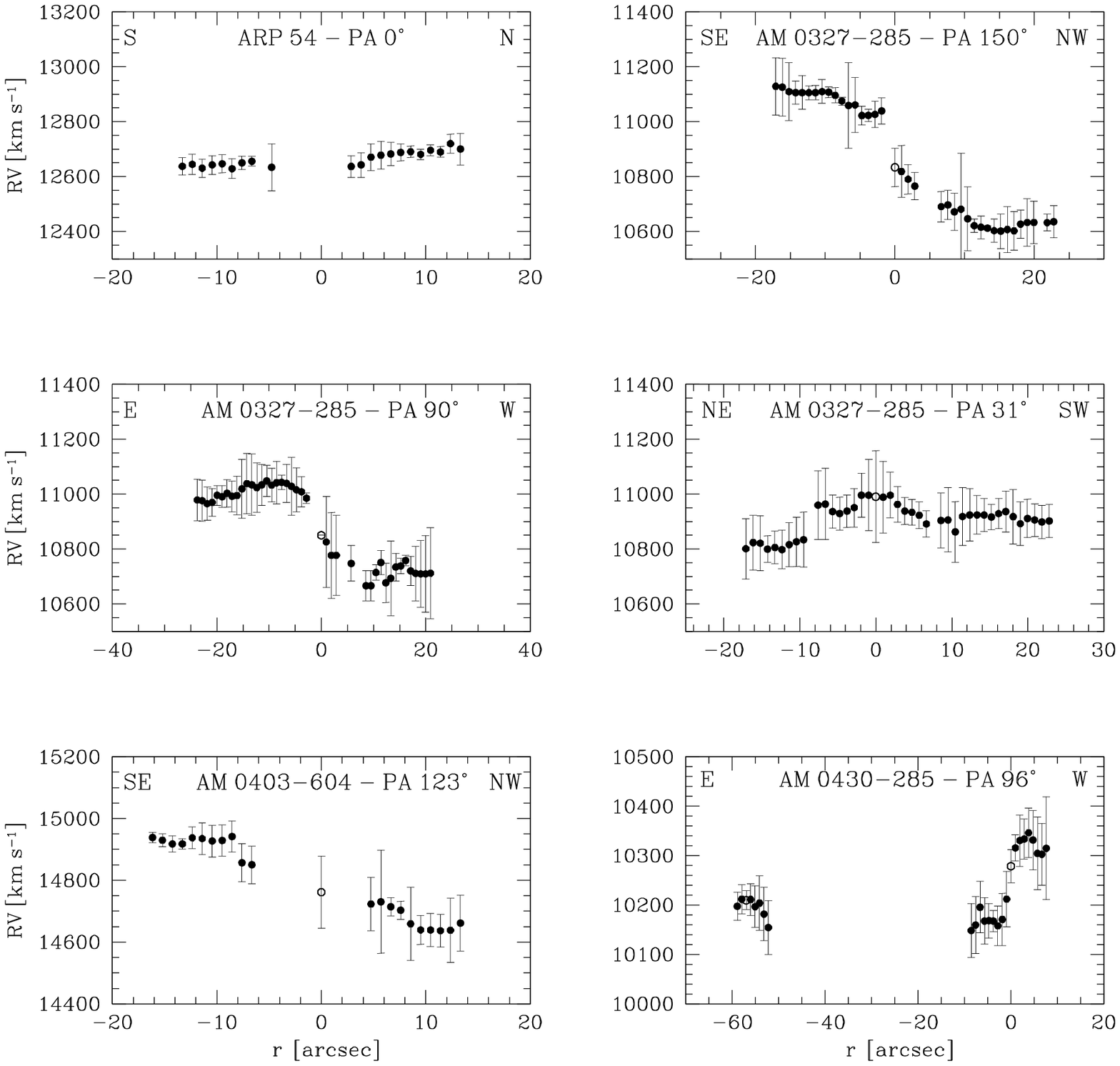}
    \caption{-- \textit{continued.}}

\end{figure*}

\addtocounter{figure}{-1}
\begin{figure*}
\centering
 \includegraphics[width=17cm]{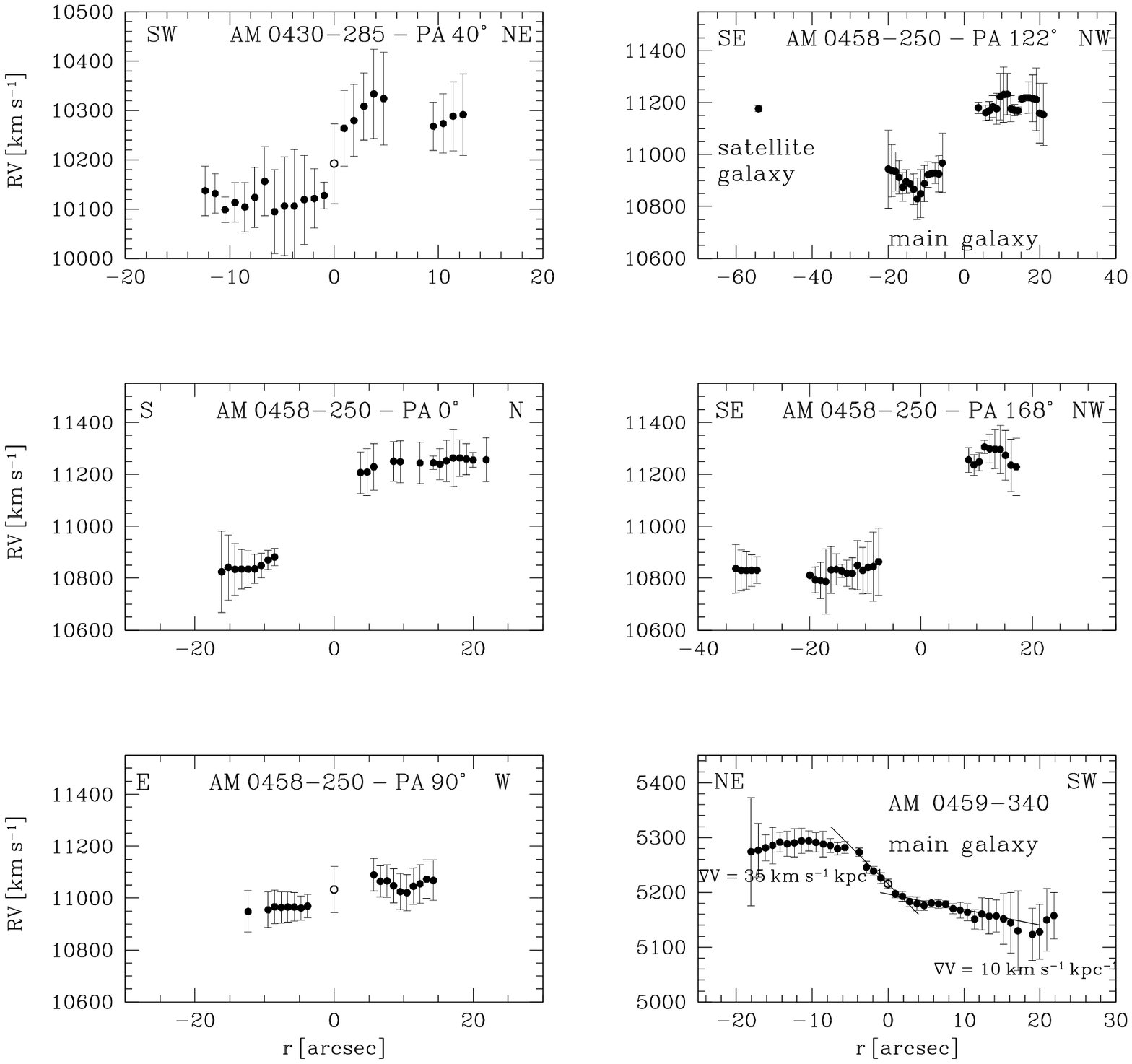}
    \caption{-- \textit{continued.}}

\end{figure*}

\addtocounter{figure}{-1}
\begin{figure*}
\centering
  \includegraphics[width=17cm]{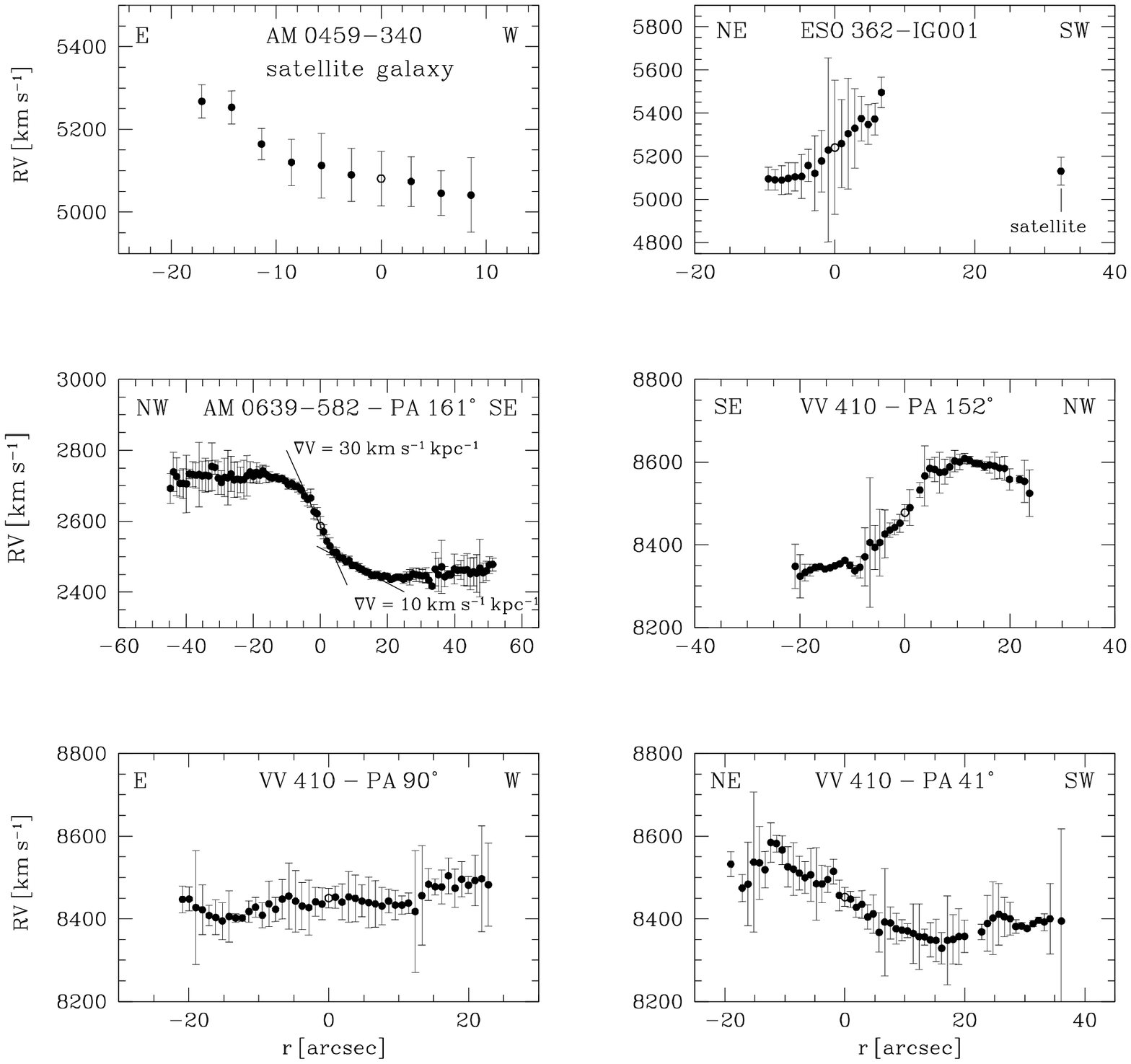}
    \caption{-- \textit{continued.}}

\end{figure*}

\addtocounter{figure}{-1}
\begin{figure*}
\centering
  \includegraphics[width=17cm]{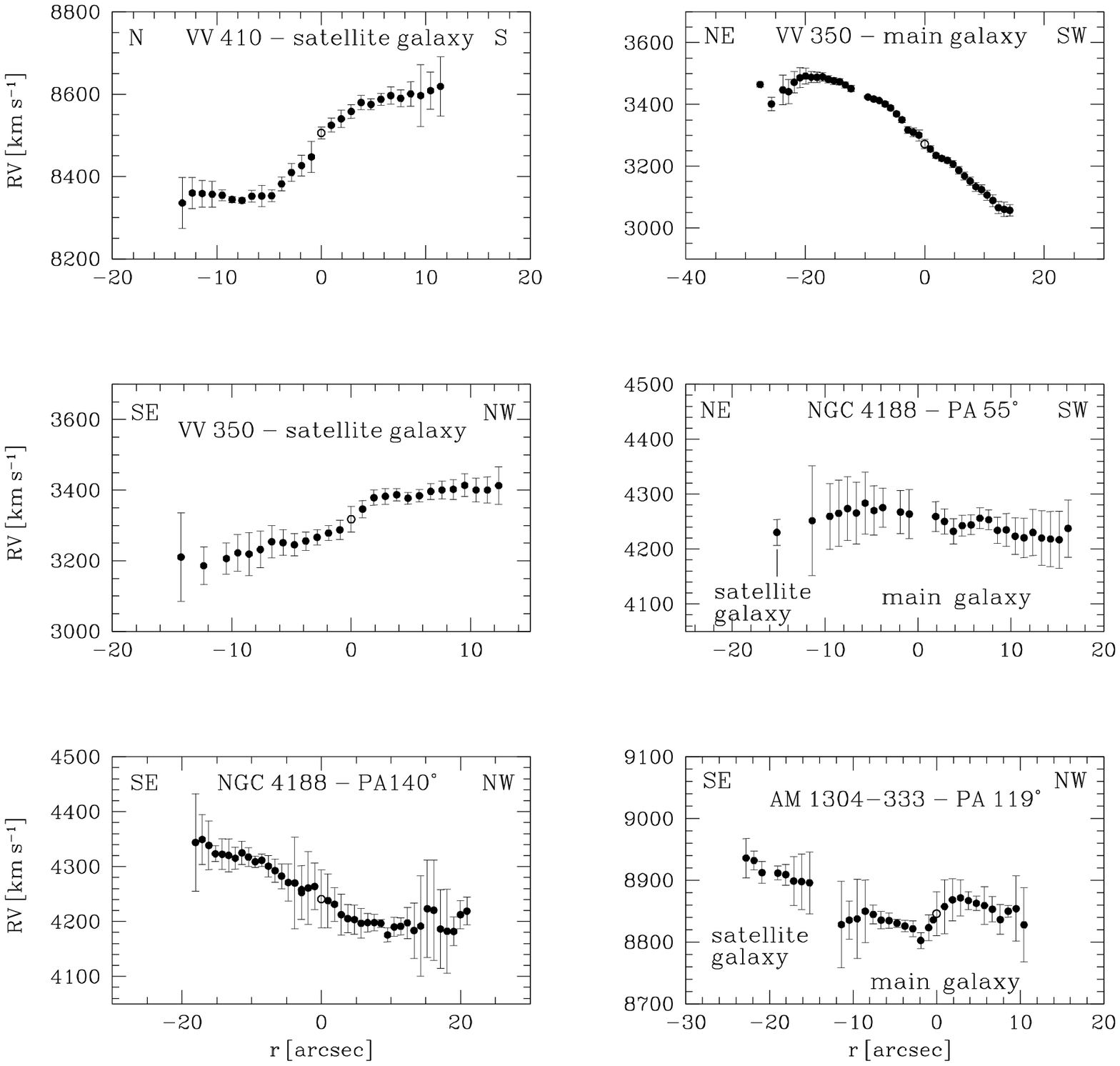}
    \caption{-- \textit{continued.}}

\end{figure*}

\addtocounter{figure}{-1}
\begin{figure*}
\centering
  \includegraphics[width=17cm]{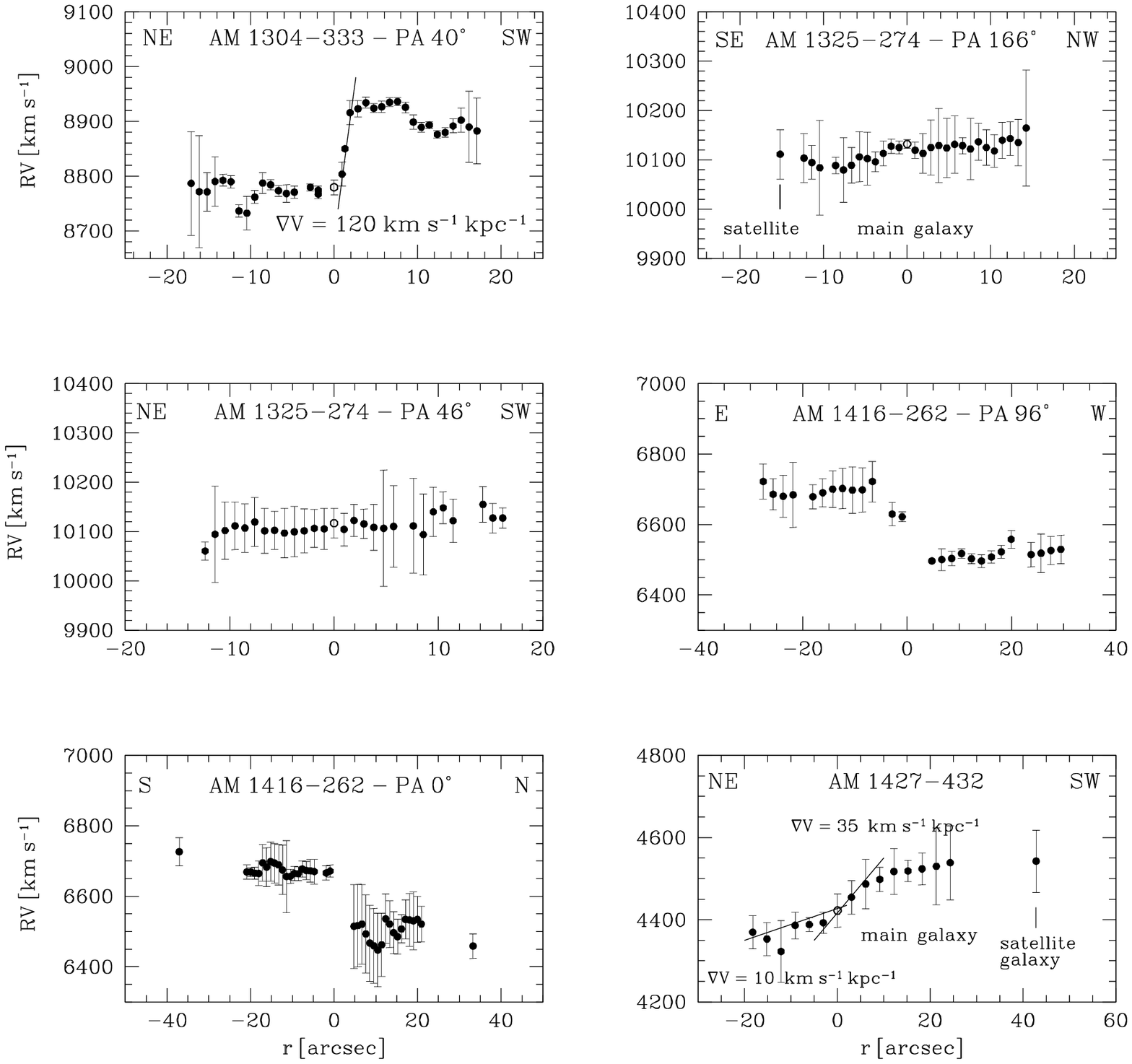}
    \caption{-- \textit{continued.}}

\end{figure*}

\addtocounter{figure}{-1}
\begin{figure*}
\centering
  \includegraphics[width=17cm]{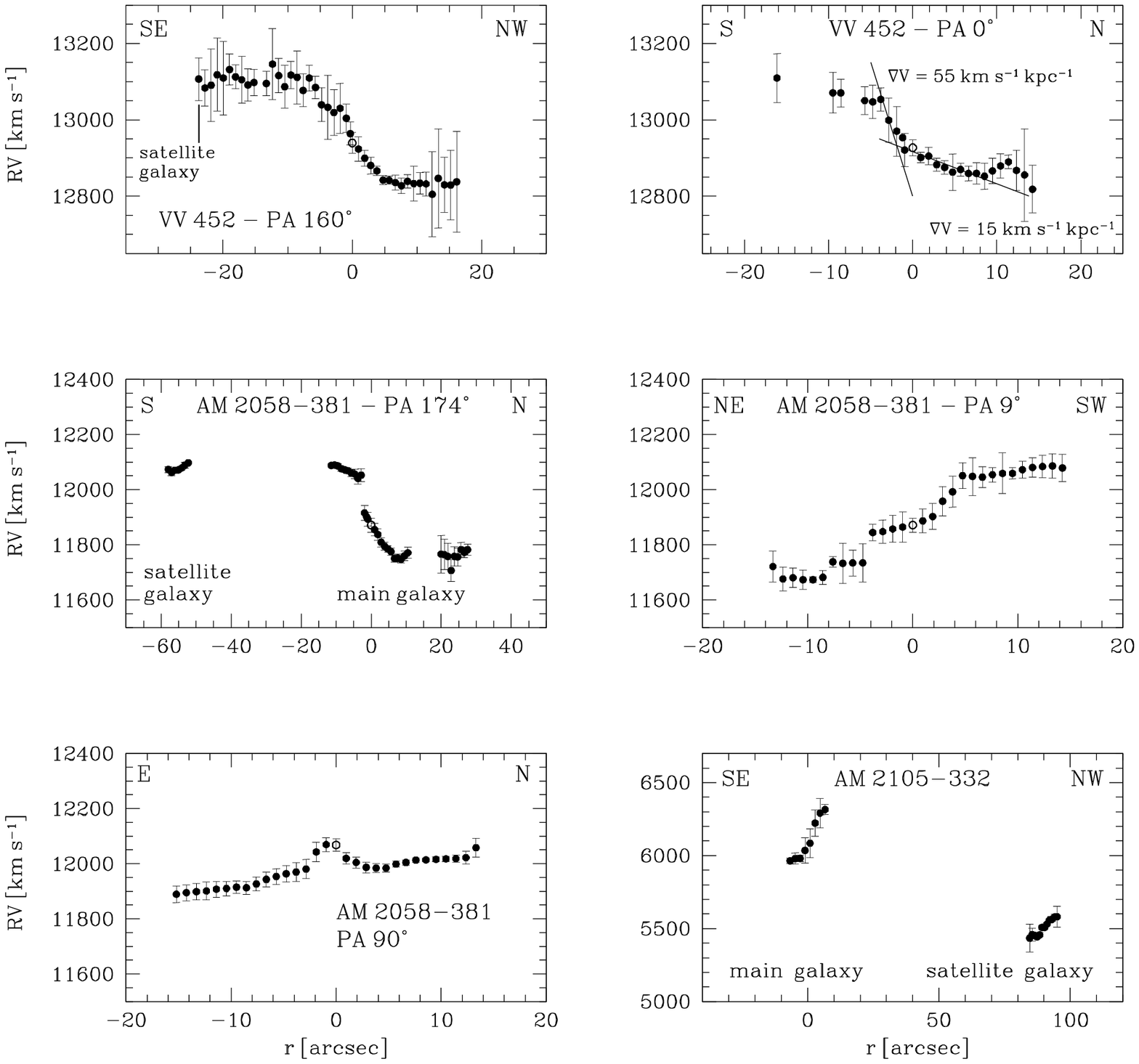}
    \caption{-- \textit{continued.}}

\end{figure*}

\addtocounter{figure}{-1}
\begin{figure*}
\centering

\includegraphics[width=17cm]{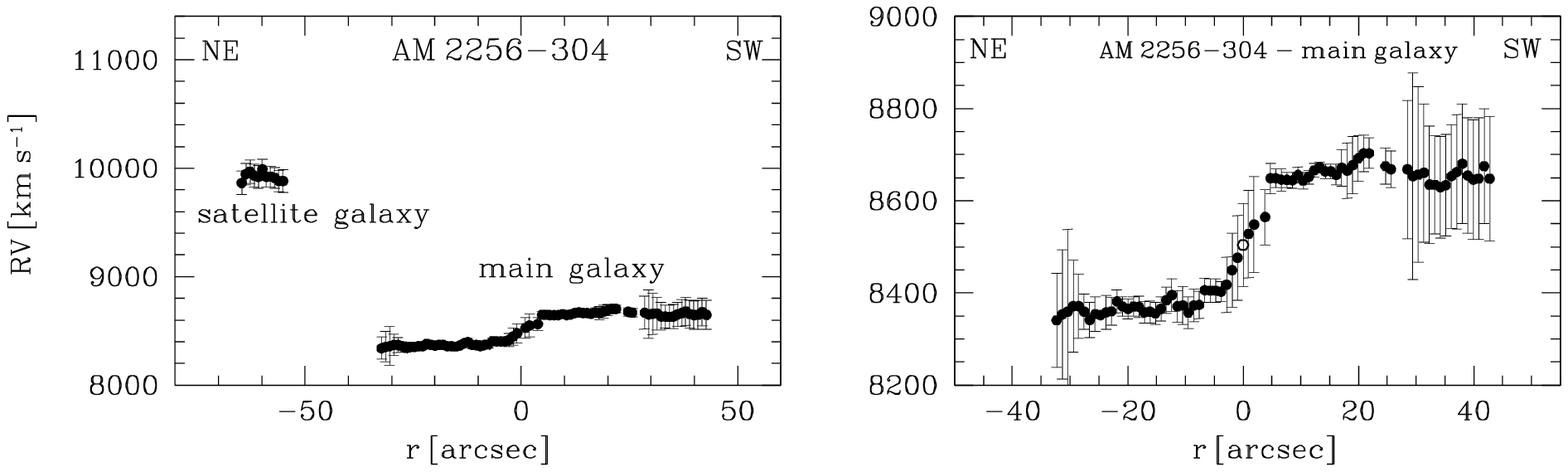}

 \caption{-- \textit{continued}}

\end{figure*}

\clearpage

\begin{figure*}
\centering

  \includegraphics[width=8cm]{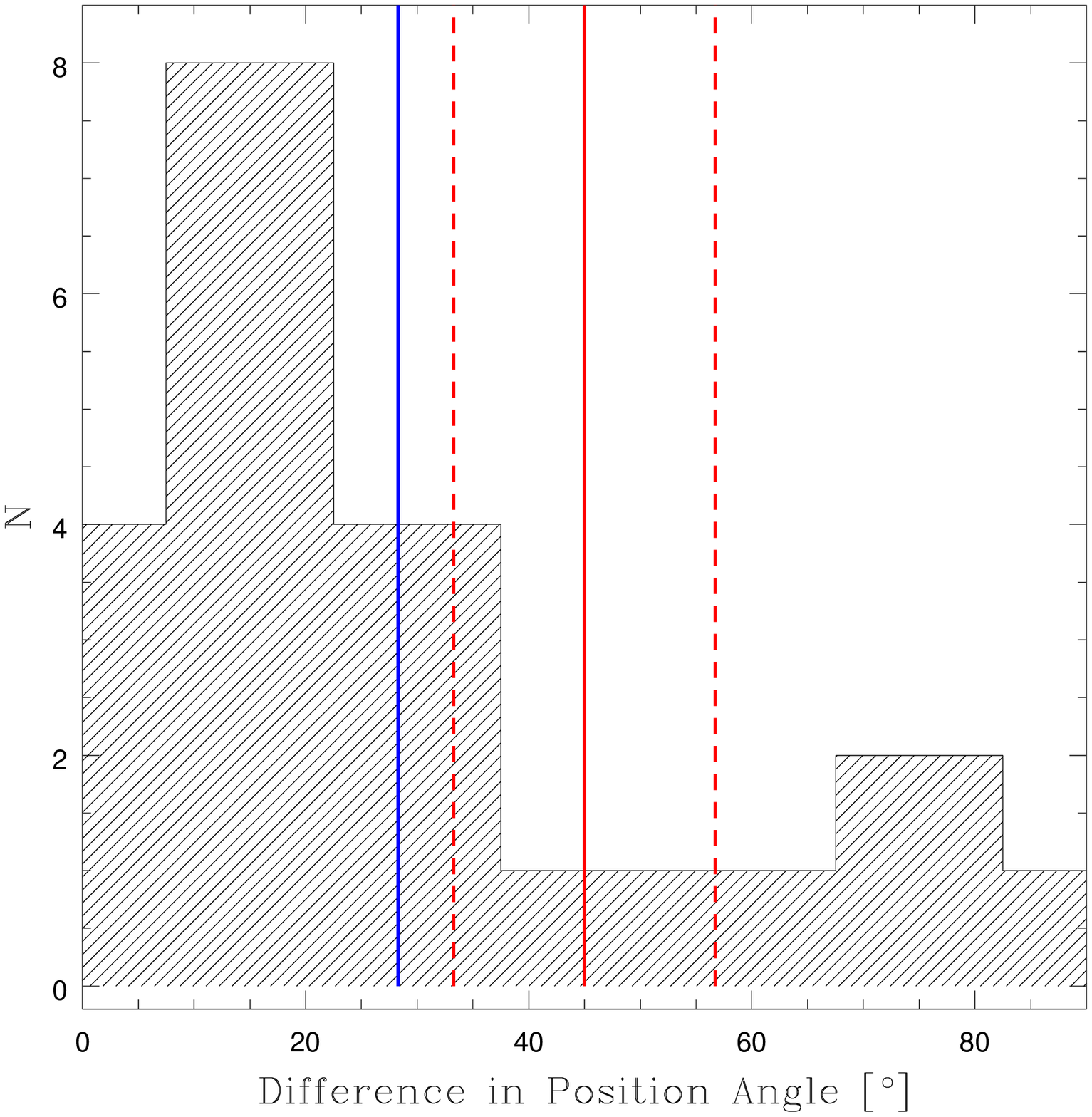}
    \caption{Histogram distribution of the difference between the position angle of the main galaxy's major axis and the position angle of the segment that connects the centers of main and satellite galaxies. There is clear excess of companions located towards the main galaxy major axis, with the observed average marked with a blue line in the plot. This indicates that the systems are in a phase of the interaction in which the orbit has a large projection on the main galaxy plane. A selection effect of M51 companions located nearer to the galaxy major axis can be discarded as there is no correlation with the system inclination. After 1500 Monte Carlo simulation runs of position angle sets, the average of the mean differences of the simulated sets (of 21 objects each) is drawn with a red solid line, and the two dashed lines correspond to the limit of 3 times the standard deviation ($\pm$\,3$\sigma$). The distribution of position angle difference in the observed sample is clearly not random.}

\end{figure*}

\clearpage

\begin{figure*}
\centering
  \includegraphics[width=8cm]{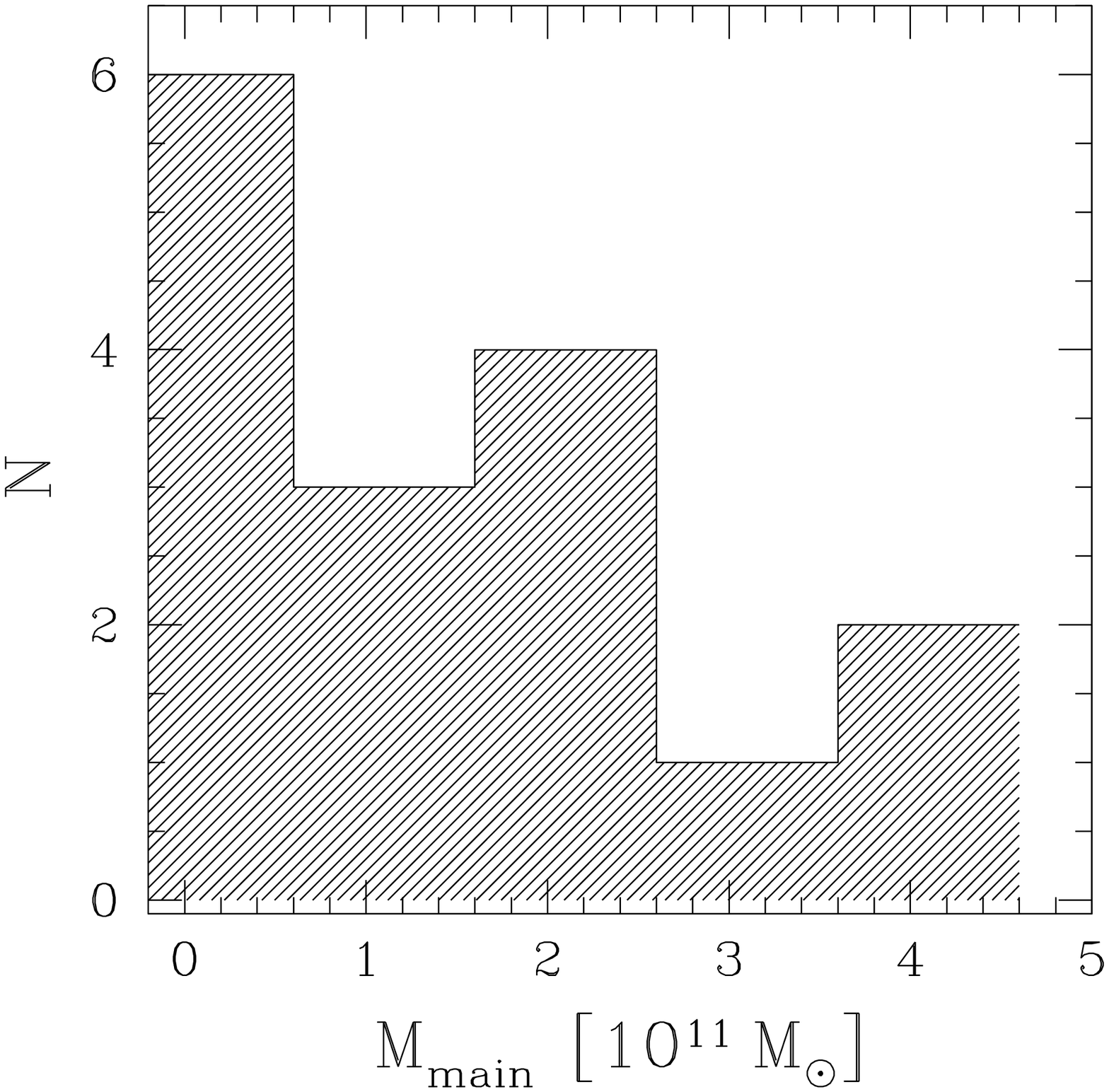}

   \includegraphics[width=8cm]{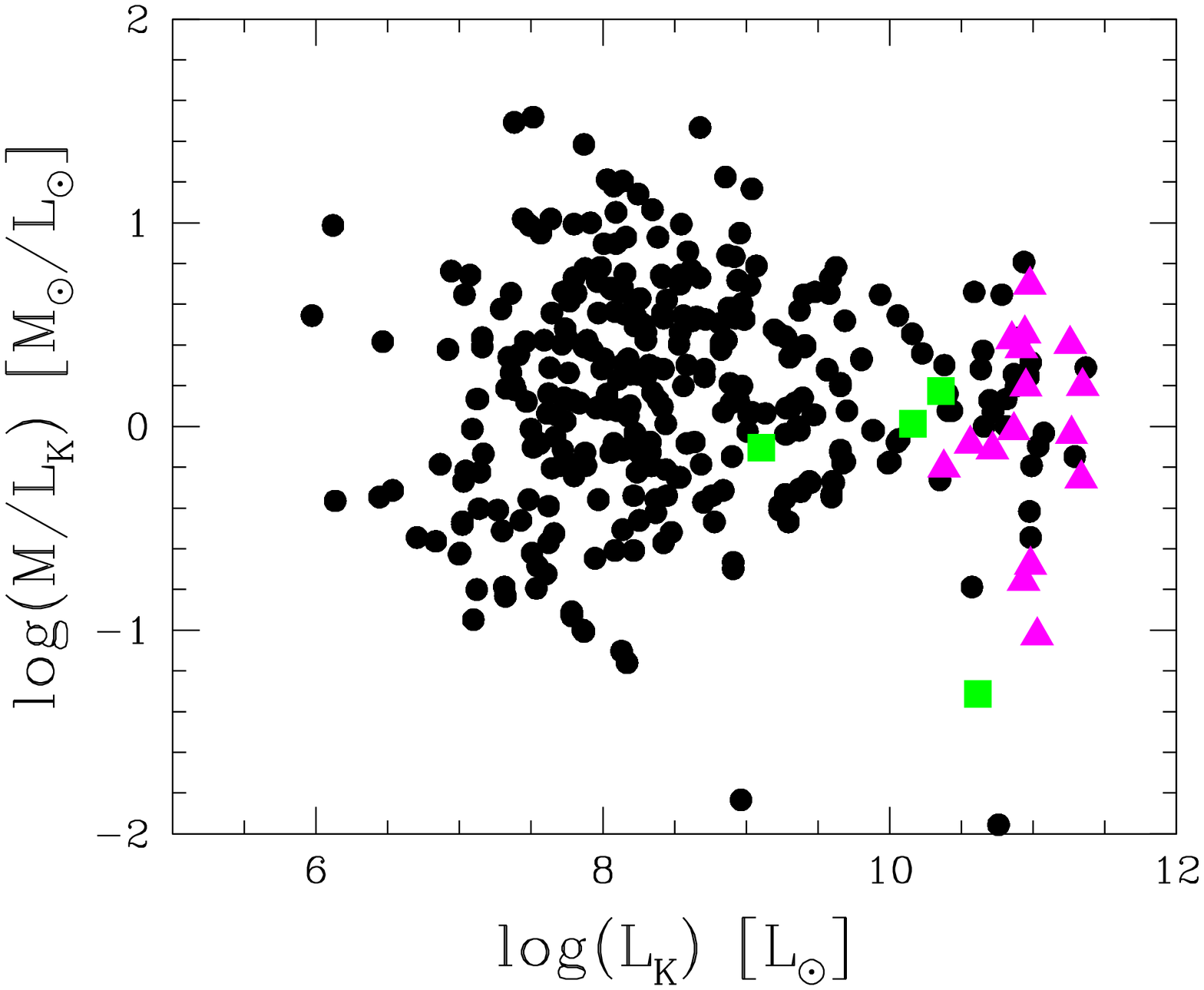}

  \caption{\textit{Up}: Histogram distribution of estimated Keplerian masses, derived for the main galaxies of the M51-type systems. \textit{Bottom}: In this log(M/L$_{K}$) vs. (L$_{K}$) plot are included the values corresponding to our sample of M51-type galaxies, magenta triangles for main galaxies and green squares for satellites galaxies, while in black circles are included the values corresponding to the galaxies of Karanchentsev et al.'s catalogue (see Karachentsev et al. 2004 and Figure\,6 of Karachentsev \& Kutkin 2005.) The K-band luminosity is in K-band solar luminosity units.}

\end{figure*}

\clearpage

\begin{figure}
 \centering
  \includegraphics[width=0.45\textwidth]{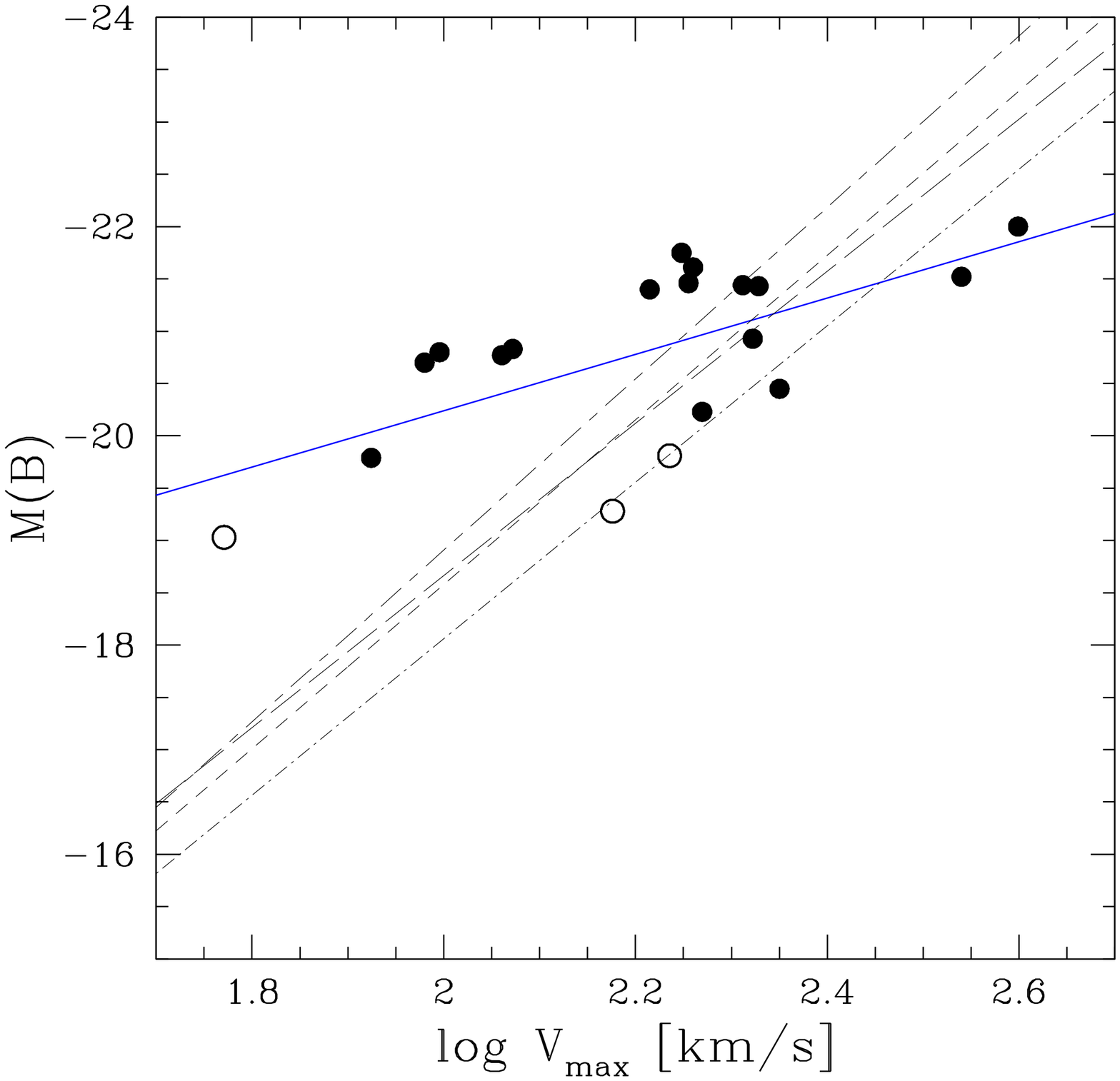}
  \includegraphics[width=0.45\textwidth]{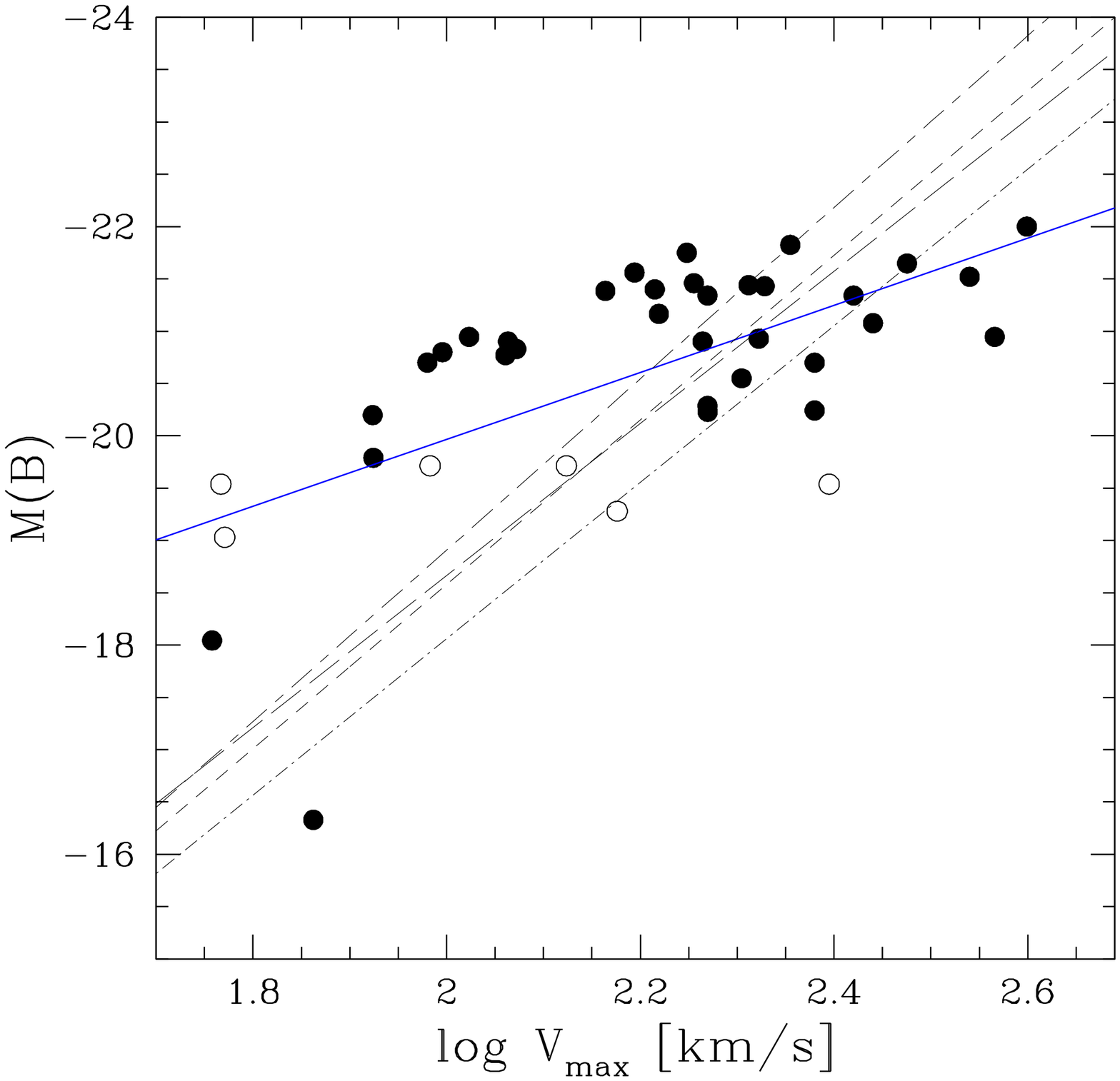}
  \includegraphics[width=0.45\textwidth]{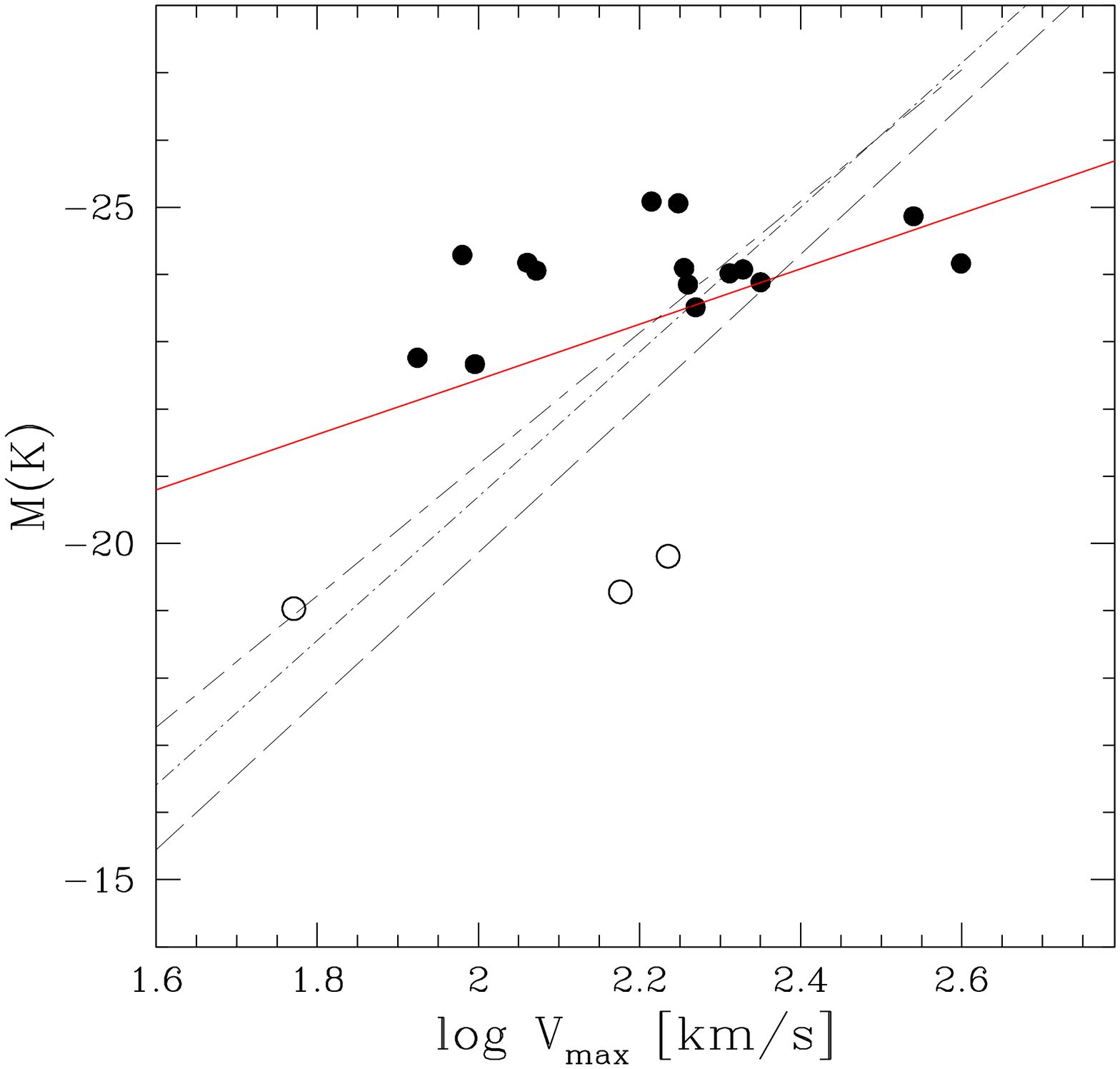}

 \caption{\footnotesize{\textit{Upper left}: Tully-Fisher relation for the sample galaxies in B-band. In \textit{blue solid line} is drawn the fitting to the data. In the figures are also plotted, for comparison, the standard Tully-Fisher relation, Tully et al. (1998) (\textit{short-dashed line}); Tully $\&$ Pierce (1998) (\textit{long-dashed line}); Pierce $\&$ Tully (1992) (\textit{dashed-dotted line}) and Kudrya \& Karachentseva (2012) (\textit{short dashed-long dashed line}). \textit{Upper right}: Tully-Fisher relation in B-band, including the galaxies of the present sample and the objects from Reshetnikov $\&$ Klimanov (2003). In \textit{blue solid line} is drawn the fitting to the data. The comparison fitting lines are the same as those plotted in figure in the upper left plot. \textit{Bottom}: K-band Tully-Fisher corresponding to the sample of the present work. In the \textit{solid line} we plot the best fitting to the data, and for comparison, we also plotted the relationships found by Kudrya \& Karachentseva (2012) (\textit{short dashed-long dashed line}), Torres-Flores et al. (2011) (\textit{long-dashed line}) and  Masters, Springob $\&$ Huchra (2008) (\textit{dashed-dotted line}). In all the plots filled circles correspond to the main galaxies while open circles to satellite galaxies.}}

\end{figure}

\clearpage

\begin{figure*}
\centering
  \includegraphics[width=15cm]{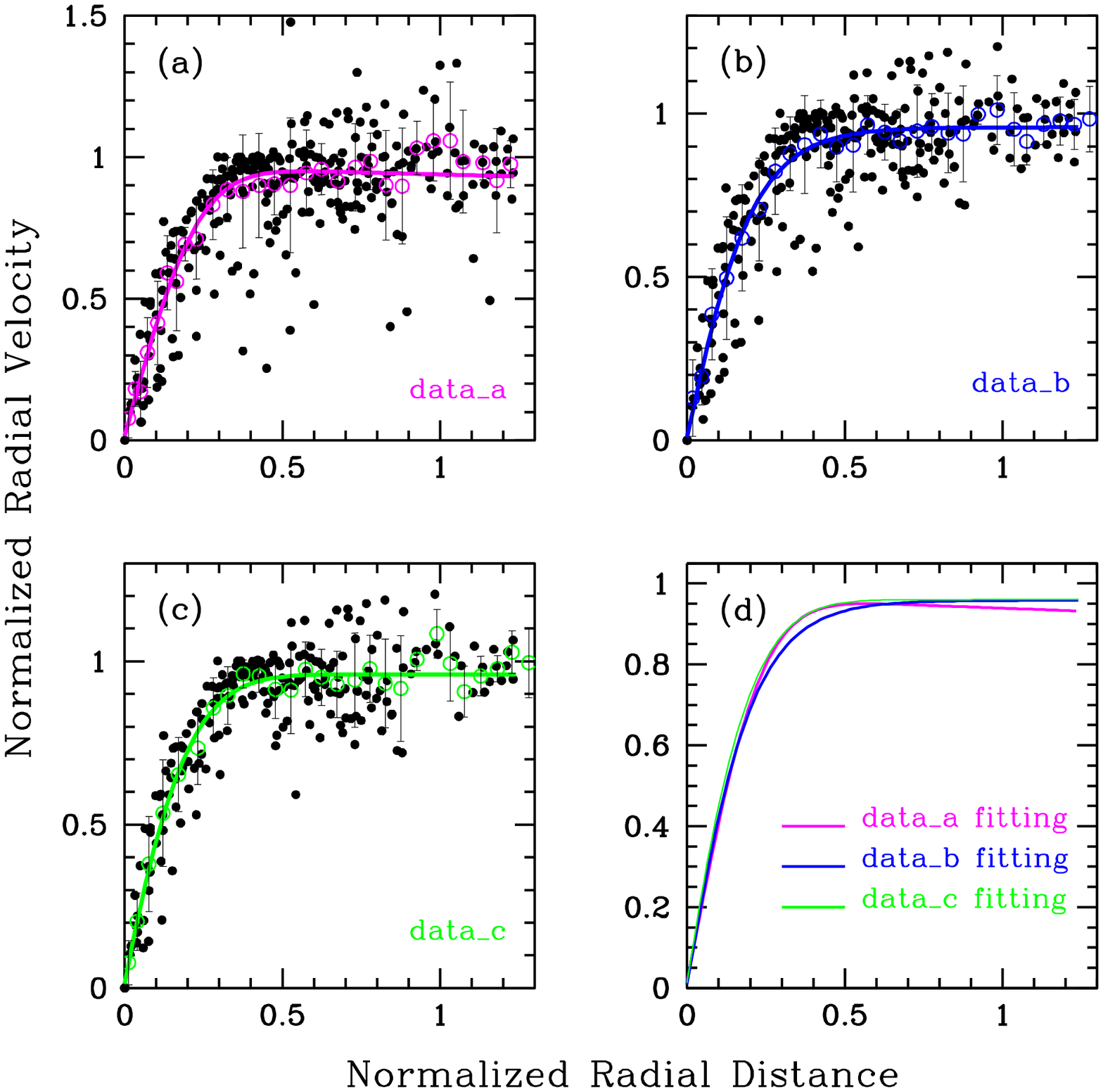}
    \caption{Panels \textit{(a)}, \textit{(b)} and \textit{(c)} show the radial velocity data with their corresponding fitting, for the three sets considered: data\_{a}, data\_{b} and data\_{c}, respectively. Panel \textit{(d)} shows the fitting for the three sets of data.}

\end{figure*}

\clearpage

\begin{figure}[!ht]
  \centering
  \includegraphics[width=0.55\textwidth]{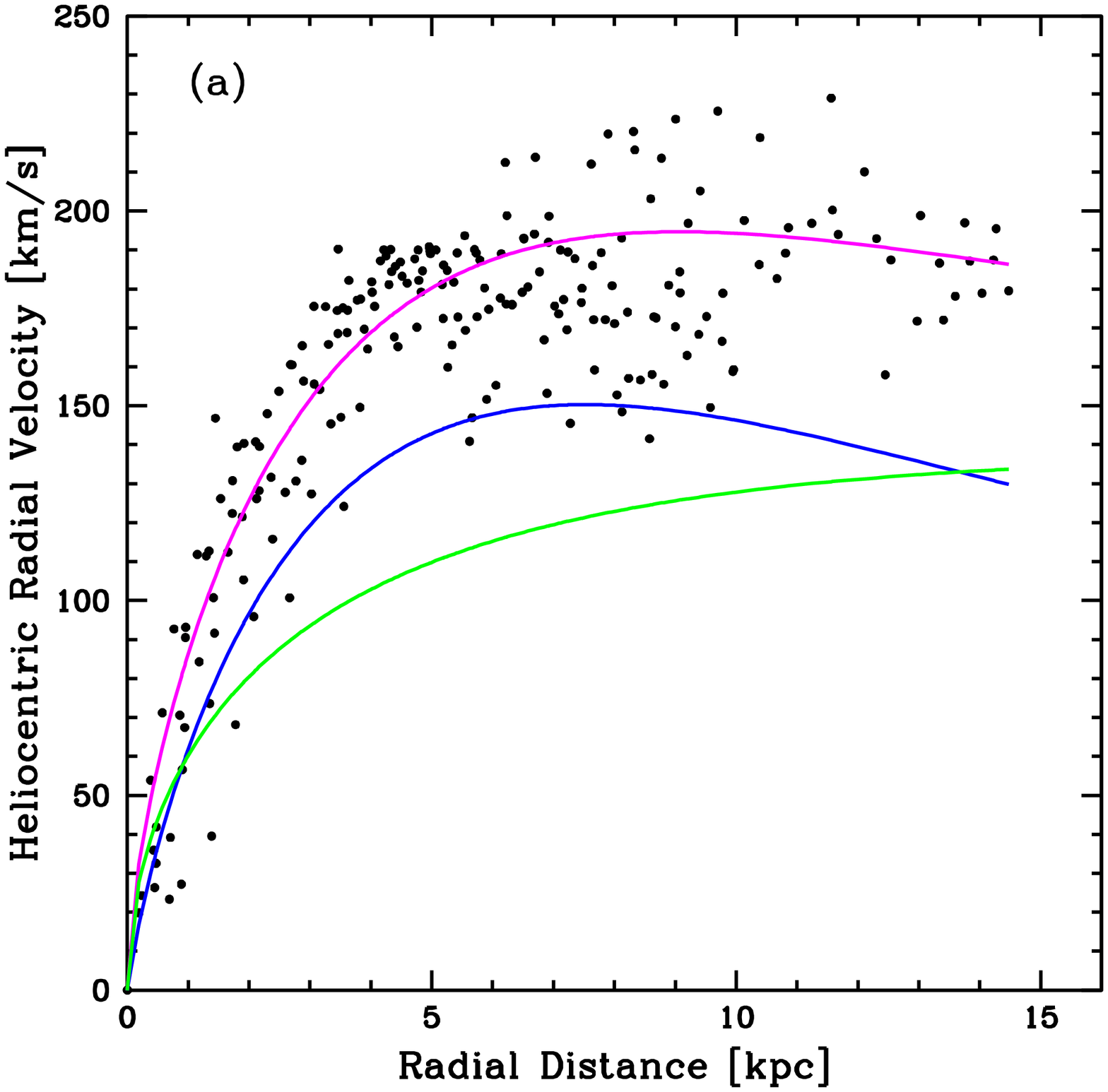}~\hfill
   \includegraphics[width=0.55\textwidth]{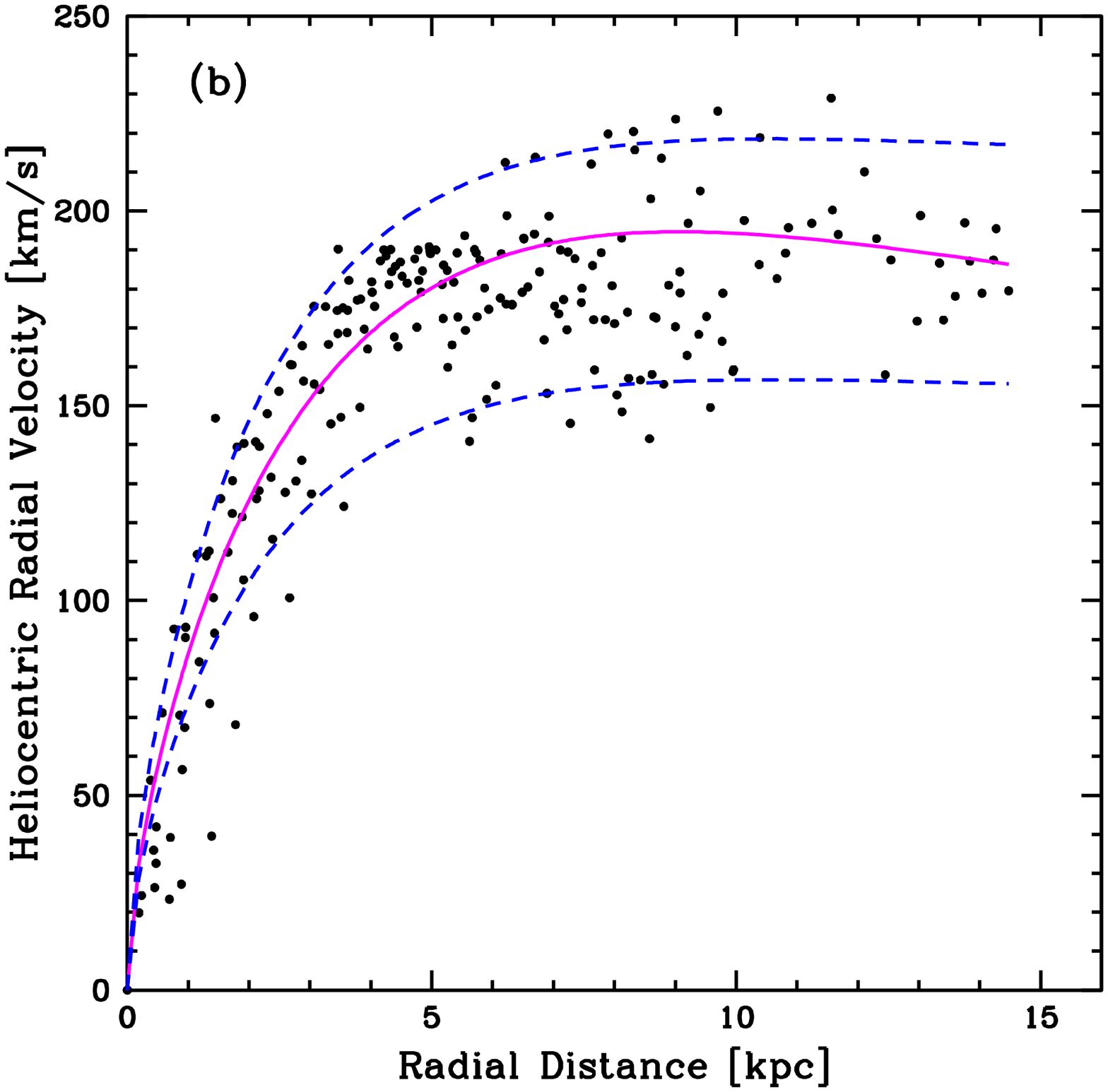}
 \includegraphics[width=0.55\textwidth]{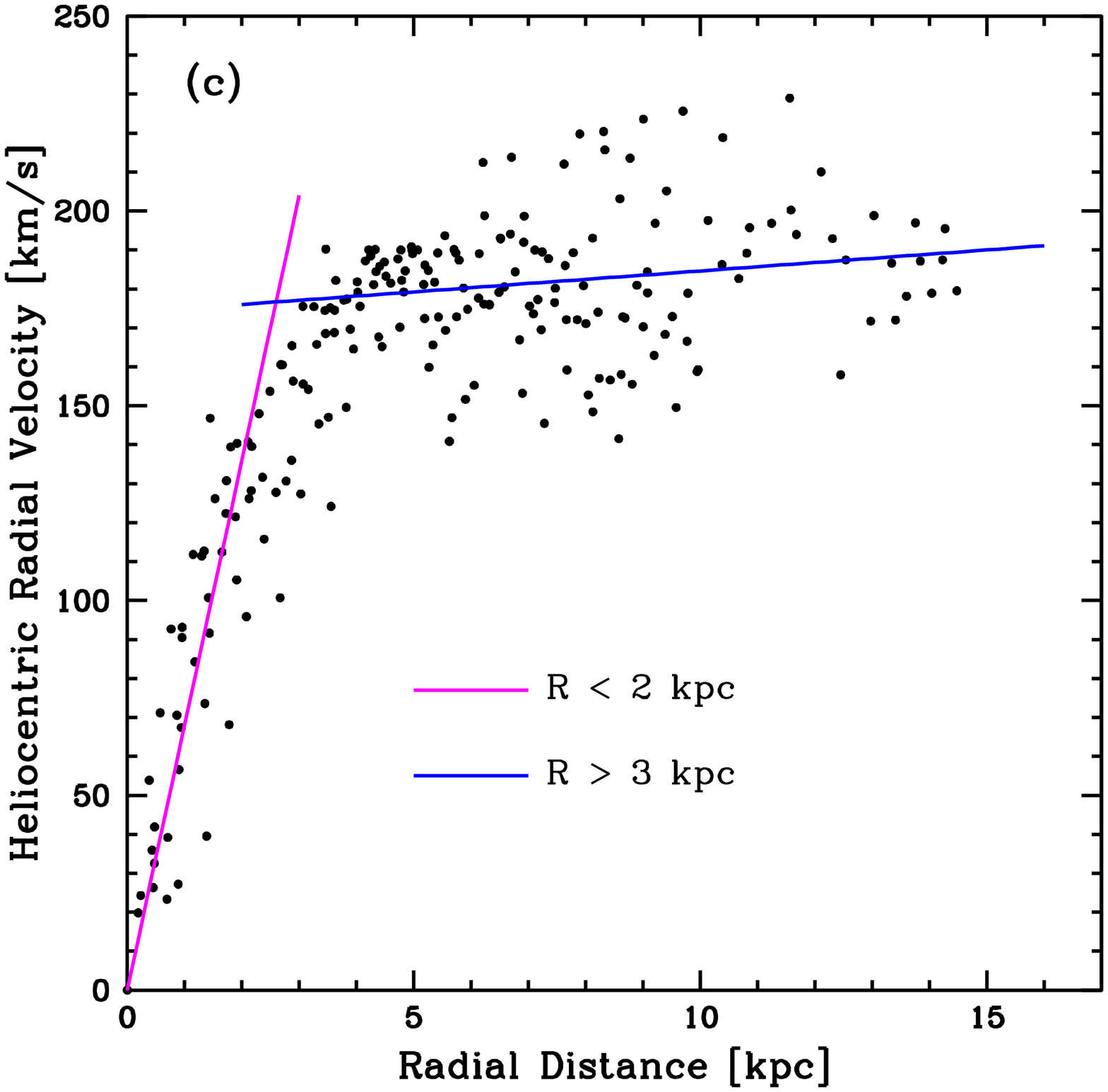}~\hfill

 \caption{Heliocentric radial velocities corresponding to set ``c".  The chosen potentials correspond to the description of Barnes $\&$ Hibbard 2009). In panel \textit{(a)} the distribution is fitted considering a potential which includes an exponential disk (Freeman 1970) in blue line, and a halo component (Navarro, Frenk $\&$ White 1996) in green line. The resulting rotation curve is plotted in magenta. In panel \textit{(b)} we plot the disk plus halo potential and two rotation curve envelopes (see text for more details). In panel \textit{(c)} we include the linear fitting for two distinctive regions of the rotation curves: the central rigid rotation region (R\,$<$\,2\,kpc) and the flat region (R\,$>$\, 3\,kpc).}
\end{figure}

\clearpage

\begin{figure*}
\centering
  \includegraphics[width=10cm]{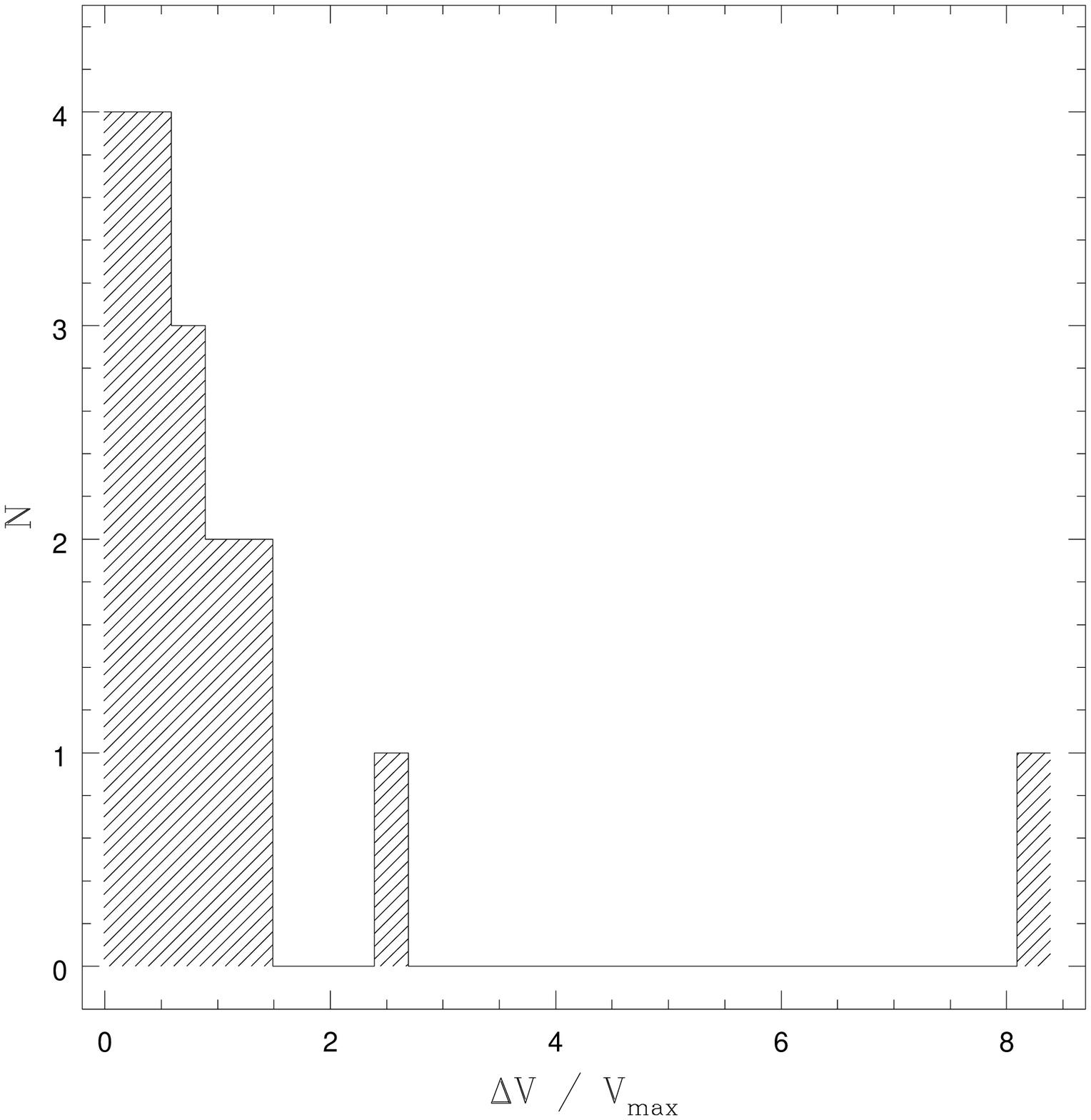}
 \caption{Distribution of radial velocity differences between main and satellite galaxies, normalized to the amplitude of the main galaxy's radial velocity curve. Most of the companions have velocities consistent with orbital motion around the main galaxy. All the systems are gravitationally bound, except for the pair AM\,2256-304, which seems to be a high velocity encounter case.}
\end{figure*}

\clearpage

\begin{table}[!ht]
\centering

\begin{tabular}{lcccccc}
\multicolumn{2}{c}
{Table 1: Object List}\\
[1ex]
\hline
\hline

Object & RA\,(J2000) & DEC\,(J2000)     &  $B_{main}$ & $B_{sat}$  & $Ks_{main}$
& $Ks_{sat}$\\
[1ex]
\hline

NGC\,633       &01 36 23.4 &  -37 19 18   &  13.33  & 15.15    & 9.92 & 10.84  \\
ARP\,54        & 02 24 02.6 &  -04 41 36   &  14.87 &   16.33   &10.91 & 13.04  \\
AM\,0327-285   & 03 29 56.1 &  -28 46 14   &  14.06&   14.55   & 11.55 & 12.78   \\
AM\,0403-604   & 04 04 27.1 &  -60 40 57   &   15.60 &   16.81   &11.46&13.01    \\
AM\,0430-285   & 04 32 11.3 &  -28 51 39       & 15.84   &    &11.52 &  12.99   \\
AM\,0458-250   & 05 00 41.9 &  -25 04 33      &  14.95 &  18.32  &10.85&13.88    \\
AM\,0459-340   & 05 01 41.2 &  -34 01 56      &  14.27 &       & 11.42 & 14.60    \\
ESO\,362-IG001   & 05 01 55.9 & -34 01 43      & 15.08 &      &  11.01 & 16.49      \\
AM\,0639-582  & 06 40 43.2 & -58 31 28    &  13.40 &  15.08     &9.07 & 11.80      \\
VV\,410    & 10 44 07.0 &  -16 28 11      &  14.50&     16.01    &11.07&12.50      \\
VV\,350         & 11 40 11.4 &  +15 20 05      &  13.93&   13.95    &9.15&10.99      \\
NGC\,4188     & 12 14 07.3 &  -12 35 10     &  14.57&        &10.89&16.68    \\
AM\,1304-333  & 13 07 08.8 &  -33 51 58     &  14.71&    16.30    &10.92&11.48      \\
AM\,1325-274  & 13 28 03.3 &  -27 55 00      &  15.10 &        &11.48&14.81     \\
AM\,1416-262  & 14 19 22.4 &  -26 38 41     &  13.83 &       &9.57&11.95      \\
AM\,1427-432  & 14 30 18.0 &  -43 33 40     &  14.96&         &10.96&     \\
VV\,452       & 16 01 16.3 &  +17 40 40      &  15.29&        &12.22&      \\
AM\,1955-170  & 19 59 29.7 &  -56 59 57   &    &   &12.13&12.77    \\
AM\,2058-381  & 21 01 39.1 &  -38 04 59     &  15.43 &&11.88&13.55      \\
AM\,2105-332  & 21 08 05.5 &  -33 13 55    &  13.90 &15.15& 9.17&11.38     \\
AM\,2256-304  & 22 58 58.6 &  -30 29 38    &   14.53 &18.25& 11.07 &14.97   \\
[1ex]
\hline

\end{tabular}
\caption{\textit{col.\,(1)}: object name; \textit{col.\,(2)}: right ascension J2000; \textit{col.\,(3)}: declination J2000; \textit{col.\,(4)}: B apparent magnitude (LEDA) of main galaxies (for the main galaxy of VV410 the value is obtained from Gunthardt 2009); \textit{col.\,(5)}: B apparent magnitude (LEDA) of satellite galaxies (for the satellite galaxy of VV410 the value is obtained from Gunthardt 2009);\textit{col.\,(6)}: K apparent magnitude (2MASS) of main galaxies; \textit{col.\,(7)}: K apparent magnitude (2MASS) of satellite galaxies.}

\end{table}

\begin{table}[!ht]
\centering

\begin{tabular}{lccccc}

\multicolumn{3}{c}
{Table 2: Systemic Heliocentric Radial Velocities.}\\
[1ex]
\hline
\hline

Object & $V_{main}$[km s$^{-1}$] & $V_{main}$ [km s$^{-1}$]     &  $V_{sat}$ [km\,s$^{-1}$] & $V_{sat}$ [km\,s$^{-1}$]  & $\delta$V [km\,s$^{-1}$] \\ [1ex]
\hline

NGC\,633      & 5098 (5) & 5103   & 5168 (10)  &        &  70      \\
ARP\,54        & 12628 (30) & 12666  &       &        &  60$^{\dag}$        \\
AM\,0327-285   & 10833 (60) & 10853  &       &        &  94$^{\ddag}$          \\
AM\,0403-604   & 14760 (50)& 14785  & 15270 &        &    510     \\
AM\,0430-285   & 10279 (30) &        & 10209 (25) &        &  70     \\
AM\,0458-250   & 11018 (20)& 11043  & 11177 (12)&        & 159           \\
AM\,0459-340   &  5215 (8) & 5209   & 5081 (30) &        & 134          \\
ESO\,362-IG001 &  5241 (50)&        & 5131 (50) &        & 110         \\
AM\,0639-582  &  2586 (20)& 2587   &  2732 (20) &        & 146         \\
VV\,410   &  8478 (20)& 8473    & 8506 (14)  &        & 28         \\
VV\,350        &  3272 (15)&        & 3318 (35)  &        &  46    \\
NGC\,4188      &  4241 (50)& 4261   & 4230 (20)  &        & 11        \\
AM\,1304-333   &  8846 (10)& 8852   & 8910 (15)  &        & 64         \\
AM\,1325-274   & 10131 (10) &        & 10111 (35)&        & 20      \\
AM\,1416-262   &       & 6600   &       &        & 139$^{\ddag}$        \\
AM\,1427-432   &  4422 (30)& 4430   & 4542 (50) &        & 120      \\
VV\,452    & 12939 (20)& 12968  & 13107 (40) &        & 168       \\
AM\,1955-170   & 17180 (80)&        &       &        &               \\
AM\,2058-381   & 11871 (20) & 11862  & 12080 (19)&        & 209       \\
AM\,2105-332   &       & 6147 (15)   &       &  5512 (7)  & 591    \\
AM\,2256-304   & 8504 (70) & 8507   & 9920 (50)  &        & 1416     \\
[1ex]
\hline

\end{tabular}

\caption{\textit{col.\,(1)}: Object name. \textit{col.\,(2)} and \textit{col.\,(4)}: Systemic Radial velocities obtained at the position of the peak of the continuum emission. \textit{col.\,(3)} and \textit{col.\,(5)}: Systemic Radial velocities obtained from the symmetry center of the radial velocity curve. \textit{col.\,6}: Modulus of the difference between the systemic radial velocities of main and satellite galaxies. Estimated uncertainties ($2\sigma$) are listed between brackets, and are derived from the empirical expression for $\sigma$ from Keel (2004). $\dag$ Radial velocity difference obtained from CASLEO spectrophotometric observations. $\ddag$ Radial velocity difference extracted from  data.}

\end{table}

\begin{table}[!ht]
\centering
\begin{tabular}{lccccccc}
\multicolumn{6}{c}
{Table 3. Kinematic and Photometric Major Axis Data for Main Galaxies.}\\
[1ex]
\hline

\footnotesize{Object} & \footnotesize{PA(KMA)[$^{\circ}$]} & \footnotesize{PA(LEDA)[$^{\circ}$]}     &  \footnotesize{$\mid$C3-C2$\mid$[$^{\circ}$]}& \footnotesize{PA(R-DSS)[$^{\circ}$]}  & \footnotesize{$\mid$C5-C2$\mid$[$^{\circ}$]} & \footnotesize{PA(Bar) [$^{\circ}$]}& \footnotesize{$\mid$C7-C2$\mid$[$^{\circ}$]}\\
[1ex]

\hline
\hline

\footnotesize{VV\,410}	& \small{175 $\pm$ 15} & \small{8} & \small{13 } &  &  & \small{152 $\pm$ 3\ } & \small{23 $\pm$ 18}\\
\footnotesize{NGC\,633}& \small{135 $\pm$ 10} &   &    &  \small{      138 $\pm$ 10  }   &  \small{ 3 $\pm$ 20 }      &     &    \\
\footnotesize{ARP\,54} & \small{130 $\pm$ 15} & \small{112 } & \small{18}&\small{123 $\pm$ 5\ \ }&  \small{ 7 $\pm$ 20 }           &              &              \\
\footnotesize{AM\,2058-381}&  \small{ 10 $\pm$ 15}     &  \small{ 2 }  & \small{ 8 }       &  \small{\ \ 8 $\pm$ 5}  &  \small{ 2 $\pm$ 20\ }      &  \small{10 $\pm$ 3}  &\small{ \ 0 $\pm$ 18 } \\
\footnotesize{AM\,0458-250} & \small{165 $\pm$ 10} & \small{167}  &  \small{2 }       & \small{168 $\pm$ 3 } &  \small{10 $\pm$ 3\ \ \  }    &  &              \\
\footnotesize{AM\,0327-285} & \small{150 $\pm$ 15} & \small{167}  & \small{17 }   &                          &               &              &              \\
\footnotesize{AM\,0430-285} & \small{170 $\pm$ 20} &  \small{79 }     & \small{81}    &  \small{50 $\pm$ 5}      &  \small{60 $\pm$ 5\ \ \ } & \small{136 $\pm$ 3\ }    & \small{34 $\pm$ 23}  \\
\footnotesize{AM\,1416-262} & \small{140 $\pm$ 20} & \small{130}  & \small{10}    &                          &              &              &              \\
\footnotesize{AM\,1304-333} &  \small{ 40 $\pm$ 20}     &              &           &          &               & \small{\ 40 $\pm$ 3 }  &  \small{\ 0 $\pm$ 23}  \\
\footnotesize{VV\,452} & \small{165 $\pm$ 20} & \small{164}  &  \small{1}       & \small{150 $\pm$ 5}  &  \small{15 $\pm$ 25\ \ }        &               &              \\
\footnotesize{NGC\,4188}  	& \small{140 $\pm$ 20} &    &   & \small{145 $\pm$ 5}  &   \small{5 $\pm$ 25\,}   &               &              \\
\footnotesize{AM\,1325-274} & \small{100 $\pm$ 25} &        &           &           &                 &               &              \\
[1ex]
\hline

\end{tabular}

\caption{\textit{col\,(1)}: Object name (those objects for which their kinematic major axis has been determined from only two position angles, are highlighted with an asterisk); \textit{col.\,(2)}: position angle of the major kinematic axis; \textit{col.\,(3)}: position angle of the major photometric axis, extracted from LEDA; \textit{col.\,(4)}: modulus of the difference between the corresponding values of \textit{columns (2)} and \textit{(3)}. \textit{col.\,(5)}: Own determinations of photometric major axis, from R-DSS images. \textit{col.\,(6)}: Modulus of the difference between values corresponding to columns \textit{(5)} and \textit{(2)}. \textit{col.\,(7)}: Position angle of the bar subsystem. \textit{col.\,(8)}: Modulus of the difference between values corresponding to columns \textit{(7)} and \textit{(2)}.}
\end{table}

\begin{table}[!ht]
\centering

\begin{tabular}{lccc}

\multicolumn{3}{c}
{Table 4: Relative positions of satellite galaxies.}\\
[1ex]
\hline
\hline
Object & major axis P.A. [$^{\circ}$]   & satellite P.A. [$^{\circ}$]&
$\Delta$(P.A.)[$^{\circ}$]\\
[1ex]
\hline

NGC\,633       & 168   & 8    & 20       \\
ARP\,54        & 90    & 76   & 14               \\
AM\,0327-285   & 150   & 150  & 0                \\
AM\,0403-604   & 117   & 63   & 54       \\
AM\,0430-285   & 103   & 99   & 4    \\
AM\,0458-250   & 158   & 123  & 35       \\
AM\,0459-340   &  20   & 162  & 38       \\
ESO\,362-IG001 &  40   & 30   & 10      \\
AM\,0639-582   &  168  & 8    &  20      \\
VV\,410        &  163  & 176  & 13        \\
VV\,350        &  51   &  48  & 3   \\
NGC\,4188      &  140  & 58   & 82      \\
AM\,1304-333   & 40   & 122   & 82       \\
AM\,1325-274   & 139  & 170   & 31  \\
AM\,1416-262   & 6    & 97   & 89       \\
AM\,1427-432   &  7   & 43   & 36   \\
VV\,452        & 150& 159  & 9     \\
AM\,1955-170   & 0  &  31      & 31 \\
AM\,2058-381   & 3 & 171  & 12       \\
AM\,2105-332   &  135     & 146  &  11          \\
AM\,2256-304   & 11 & 11   &    0       \\
[1ex]
\hline

\end{tabular}

\caption{\textit{col.\,(1)}: Object name; \textit{col.\,(2)}: position angle of the photometric major axis of the main galaxy; \textit{col.\,(3)}: position angle of the center of the satellite galaxy, as measured from the center of the main galaxy; \textit{col.\,(4)}: modulus of the position angle difference between the values of \textit{col.\,(2)} and \textit{col.\,(3)}.}

\end{table}

\begin{table}[!ht]
\centering

\begin{tabular}{lccccc}

\multicolumn{3}{c}
{Table 5: Keplerian mass estimations.}\\
[1ex]
\hline
\hline
Object & inclination\,[$^{\circ}$] & PA\,(OBS)\,[$^{\circ}$]&
 mayor axis PA\,[$^{\circ}$]& R\,[kpc] & Mass\,[10$^{11}\,$M$_{\odot}$]\\
[1ex]
\hline

AM\,0639-582 		& 65  & 161 & 174 & 8  & 0.40  \,$\pm$\,0.05    \\
VV\,410\,(A)		& 51  & 152 & 170 & 14 & 1.4   \,$\pm$\,0.2     \\
VV\,410\,(B)		& 68  & 0   &   0 &  7 & 0.34  \,$\pm$\,0.06   	\\
NGC\,633\,(A)		& 49  & 136 & 136 & 6  & 0.20  \,$\pm$\,0.09   	\\
NGC\,633\,(B) 		& 65  & 0   &   0 & 2  & 0.020 \,$\pm$\,0.009 	\\
ARP\,54     		& 53  & 90  & 90  & 25 & 3.5   \,$\pm$\,0.2     \\
AM\,2256-304		& 58  & 11  & 11  & 25 & 2.5   \,$\pm$\,0.3     \\
AM\,2058-381		& 60  & 9   & 9   & 11 & 2.0   \,$\pm$\,0.2     \\
AM\,0459-340\,(A)  & 65  & 20  & 28  & 6  & 0.15  \,$\pm$\,0.09    \\
AM\,0459-340\,(B)  & 63  & 0   &  0  & 6  & 0.010 \,$\pm$\,0.005  	\\
AM\,0458-250		& 43  & 168 & 152 & 16  & 4.6  \,$\pm$\,2.0   	\\
AM\,0430-285  		& 50  & 40  & 40  & 10  & 0.15 \,$\pm$\,0.09    \\
AM\,1416-262  		& 45  & 0   & 6   & 16  & 1.2  \,$\pm$\,0.3     \\
AM\,0403-604 		& 46  & 123 & 117 & 15  & 1.7  \,$\pm$\,0.3     \\
AM\,1304-333 		& 60  & 40  & 40  & 10  & 0.10 \,$\pm$\,0.09    \\
ESO\,362-IG001     & 51  & 29  & 29  & 3   & 0.30 \,$\pm$\,0.03    \\
VV\,452 			& 50  & 160 & 150 & 20  & 1.9  \,$\pm$\,0.3     \\
VV\,350\,(A)		& 73  & 55  & 55  & 6   & 0.7  \,$\pm$\,0.1   	\\
VV\,350\,(B)		& 66  & 126 & 96  & 4   & 0.15 \,$\pm$\,0.03  	\\
AM\,0327-285  		& 44  & 150 & 150 & 14  & 4.7  \,$\pm$\,2.0       \\
[1ex]
\hline

\end{tabular}

\caption{\textit{col.\,(1)}: Object name; \textit{col.\,(2)}: main galaxy inclination (angle between the disk plane and the line of sight); \textit{col.\,(3)}: observed position angle; \textit{col.\,(4)}: major axis position angle; \textit{col.\,(5)}: the listed value correspond to the outermost position for which a spectroscopic measurement have been made; \textit{col.\,(6)}: Keplerian masses (corrected for inclination of the disk and deviation of the observed position angle with respect to the direction of the major axis.}

\end{table}

\begin{table}[!ht]
\centering

\begin{tabular}{lccc}

\multicolumn{3}{c}
{Table 6: Data used for deriving the synthetic curve.}\\
[1ex]
\hline
\hline
Object &data\_{a}&data\_{b}& data\_{c}\\
[1ex]
\hline

NGC\,633       & \checkmark &   \checkmark & \checkmark       \\
ARP\,54        & \checkmark & \checkmark & \checkmark       \\
AM\,0327-285   & \checkmark & \checkmark & ...     \\
AM\,0403-604   & \checkmark & ... & ...    \\
AM\,0458-250   & \checkmark & \checkmark & \checkmark     \\
AM\,0459-340   & \checkmark & \checkmark & \checkmark   \\
AM\,0639-582   & \checkmark & \checkmark & \checkmark  \\
VV\,410        & \checkmark & \checkmark & ...      \\
VV\,350        & \checkmark & \checkmark & \checkmark \\
NGC\,4188      & \checkmark & ...   & ...     \\
VV\,452        & \checkmark & \checkmark & \checkmark  \\
AM\,2058-381   & \checkmark & \checkmark & \checkmark    \\
AM\,2256-304   & \checkmark & \checkmark & \checkmark \\
[1ex]
\hline

\end{tabular}

\caption{\textit{col.\,(1)}: object designation; \textit{cols.\,(2)} to \textit{(4)}: selected objects used for three data sets, data\_{a}, data\_{b} and data\_{c}, from which synthetic curves are derived.}

\end{table}

\clearpage

\begin{landscape}

\begin{table}[!ht]
\centering

\begin{tabular}{lcccccccccccccc}

\multicolumn{7}{c}
{Table 7: Coefficients of the synthetic rotation curve fittings.}\\
[1ex]
\hline
\hline
 & \footnotesize{function} & \footnotesize{a} & \footnotesize{b} & \footnotesize{c} & \footnotesize{d} & \footnotesize{e} & \footnotesize{f} & \footnotesize{$r^{2}$} & \footnotesize{${\sigma}$} & \scriptsize{M$_{d}$ [10$^{10}$\,M$_{\odot}$]} & \footnotesize{$\textit{a}_{d}$} \scriptsize{[kpc]} & \scriptsize{$M_{h}$ [10$^{10}$\,M$_{\odot}$]} & \footnotesize{$\textit{a}_{h}$} \scriptsize{[kpc]} & \footnotesize{$\sigma$ [km\,s$^{-1}$]}\\
[1ex]
\hline

\scriptsize{data\_{a}} & \scriptsize{ADC}      & \scriptsize{3.673} & \scriptsize{0.051} & \scriptsize{0.086} & \scriptsize{0.179}  & \scriptsize{50.410} & \scriptsize{-0.871} & \scriptsize{0.76} & \scriptsize{0.15}\\
\scriptsize{data\_{b}} & \scriptsize{GaussMod} & \scriptsize{1.508} & \scriptsize{8641.187} & \scriptsize{8641.1024} & \scriptsize{70015.656} & \scriptsize{-0.549} && \scriptsize{0.82}  & \scriptsize{0.12}\\
\scriptsize{data\_{c}} & \scriptsize{SDC} & \scriptsize{3.287} & \scriptsize{1.690} & \scriptsize{3.625} & \scriptsize{0.220} & \scriptsize{-2.327}	&& \scriptsize{0.85}  & \scriptsize{0.10} & \scriptsize{4.75\,$\pm$\,0.35} & \scriptsize{3.5\,$\pm$\,0.2} & \scriptsize{4.0\,$\pm$\,1.1} & \scriptsize{10.5\,$\pm$\,0.3} & \scriptsize{19}\\
[1ex]
\hline

\end{tabular}

\caption{\textit{col.\,(1)}: data set used for the fitting function; \textit{col.\,(2)}: fitting function denomination;     \textit{cols.\,(3)} to \textit{(8)}: coefficients `a', `b', `c', `d', `e' and `f', of the corresponding fitting functions, which are listed in the Appendix; \textit{col.\,(9)}: Coefficient of determination; \textit{col.\,(10)}: fitting standard deviation; \textit{col.\,(11)} : disk mass derived by using the potentials proposed by Barnes $\&$ Hibbard (2009) simulation; \textit{col.\,(12)}: disk scale radius; \textit{col.\,(14)}: halo mass; \textit{col.\,(14)}: halo scale radius; \textit{col.\,(15)}: fitting standard deviation.}

\end{table}

\end{landscape}

\end{document}